\documentclass[preprint, aip]{revtex4-2}

\usepackage[ruled, vlined, nosemicolon]{algorithm2e} 

\usepackage{amsmath,amssymb}           
\usepackage{graphicx}           
\usepackage{hyperref}           
\usepackage{siunitx}            
\usepackage{comment}
\usepackage{bm}                 
\usepackage[colorinlistoftodos]{todonotes}
\usepackage{tikz}
\usetikzlibrary{arrows.meta}
\tikzset{
    graphnode/.style={
        circle,
        draw,
        minimum size=8mm,
        font=\small
    },
    graphedge/.style={
        ->,
        >=Stealth,
        line width=0.8pt
    }
}

\begin{document}


\title{Zermelo's navigation problem through the lens of quantum annealing: \\ How the Landau-Zener approximation leads to an efficient classical solution}

\author{Sølve Selstø}
\affiliation{Faculty of Technology, Art and Design, Oslo Metropolitan University, NO-0130 Oslo, Norway}

\author{Tor Kristian Dahle}
\affiliation{Faculty of Technology, Art and Design, Oslo Metropolitan University, NO-0130 Oslo, Norway}

\author{Sergiy Denysov}
\affiliation{Faculty of Technology, Art and Design, Oslo Metropolitan University, NO-0130 Oslo, Norway}

\author{Yves-Laurent Ariel Rezus}
\affiliation{Amsterdam University of Applied Sciences (AUAS), Faculty of Technology, Amsterdam, the Netherlands}

\author{Leiv Øyehaug}
\affiliation{Faculty of Technology, Art and Design, Oslo Metropolitan University, NO-0130 Oslo, Norway}

\date{\today}

\begin{abstract}
The river-crossing problem, also known as Zermelo’s navigation problem, is a classic example of an optimization problem with practical relevance and a scalable degree of complexity. It asks for the optimal trajectory of a vessel moving through a water flow field and provides a setting in which physics, variational methods, and optimization are naturally intertwined. 
We state a version of Zermelo’s problem and then solve it as formulate it as an adiabatic quantum-computing problem using quantum trits, or qutrits for short. 
The construction includes a penalty term that enforces the prescribed boundary conditions and an exploration term that allows the system to move through intermediate configurations toward the optimal feasible path. In the adiabatic description, the evolution proceeds through a sequence of avoided crossings, so that the resulting fidelity can be estimated using the Landau-Zener formula. Remarkably, the regime in which this approximation is valid also provides a deterministic way to identify the correct solution with computational effort that scales only quadratically with the problem size. Thus, a quantum formulation initially motivated by the apparent exponential complexity of the problem reveals an underlying classical structure that can be exploited efficiently. Our approach also provides a pedagogical illustration of how a real-world optimization problem can be cast into a quantum-annealing framework and then analyzed using the Schrödinger equation, avoided crossings, and Landau-Zener theory.
\end{abstract}

\maketitle

\section{Introduction}
\label{sec:Intro}

The impact the emerging technology of quantum computing will have on society is hard to assess. It is no exaggeration, however, to say that many informed voices predicts it to be huge. This is not only manifested by huge investments in tech and public funding, but also by enormous efforts in academic research -- both when it comes to hardware and software developments. It is crucial that we try and prepare for the changes that these developments are expected to impose on society and professional life~\cite{Fox2020}. This, of course, also applies to education. 

Quantum computing may be divided into two main directions: Digtal or gate-based quantum computing and analogue quantum computing. Despite the fact that the world's first commercial quantum computer, the D-Wave, is a \textit{quantum annealer} belonging to the latter category, the gate-based is by far the dominating framework in literature. 

Adiabatic quantum computing and the closely related notion of quantum annealing aims to solve optimization problems by encoding the solution into the ground state of a Hamiltonian in a controlled manner~\cite{McGeoch2014}. Since many real-life problems boil down to optimization, such an analogue quantum-approach to problem solving may be quite significant if it proves able to solve problems beyond the feasibility for classical computers. Such problems are virtually omnipresent within, e.g., logistics, machine learning, modeling and predicting -- not to mention quantum physics and chemistry itself.

Although adiabatic quantum computing is formally equivalent to universal, gate-based quantum computing mathematically, it does represent an approach which is conceptually different, one that possibly could guide intuition and imagination towards novel quantum schemes for solving problems. Thus, it should hold a prominent place within STEM education -- alongside the dominating realm of gate-based quantum computing.

In this work we present a specific problem and outline how it may be formulated and solved by means of adiabatic quantum computing. 

\section{The Problem Formulation}
\label{sec:Problem}

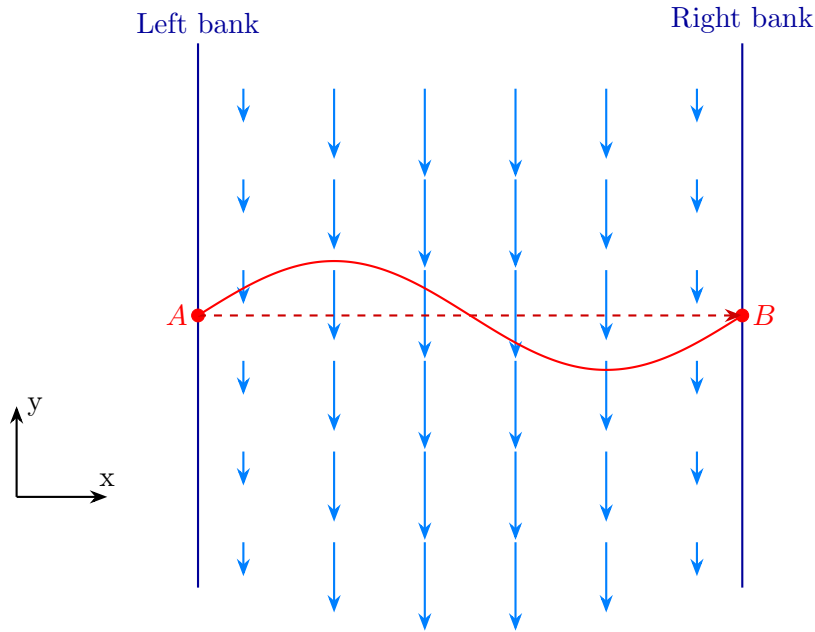
\begin{figure}[h!]
    \centering
    \begin{tikzpicture}[scale=1.2, >=Stealth]
        \draw[thick, blue!60!black] (0,0) -- (0,6);
        \draw[thick, blue!60!black] (6,0) -- (6,6);
        \node[blue!60!black, above] at (0,6) {Left bank};
        \node[blue!60!black, above] at (6,6) {Right bank};

        \filldraw[red] (0,3) circle (2pt) node[left] {$A$};
        \filldraw[red] (6,3) circle (2pt) node[right] {$B$};

        \foreach \y in {0.5,1.5,...,5.5} {
            \foreach \x in {0.5,1.5,...,5.5} {
                \pgfmathsetmacro{\d}{abs(\x - 3.0)}
                \pgfmathsetmacro{\scale}{1.0 - 0.1 * \d * \d}
                \draw[->, thick, blue!50!cyan] (\x,\y) -- ++(0,-\scale);
            }
        }

        \draw[->, thick, black] (-2,1) -- (-2,2) node[above, right] {y};
        \draw[->, thick, black] (-2,1) -- (-1,1) node[right, above] {x};

        \draw[->, thick, red!80!black, dashed] (0,3) -- (6,3);
        \draw[thick, red, domain=0:6, samples=100, smooth, variable=\x]
            plot ({\x}, {3 + 0.6*sin(deg(2*pi*\x/6))});
    \end{tikzpicture}
    \caption{Illustration of an smooth path across the river from point $A$ at the left bank to $B$ at the right bank. The current is stronger in the middle of the river. The current function, illustrated by arrows pointing downwards, depends only on the distance $x$ from the left bank.}
    \label{fig:RiverCrossingCont}
\end{figure}

Zermelo's navigation problem can be formulated in the following way~\cite{Zermelo1931}: Suppose a vessel with a fixed velocity relative to water is to cross a wide river with fixed starting and ending points. Given the river current as a function of position, what is the optimal path across the river? 
By ``optimal'' we will mean the route that minimizes the time spent and, thus, also the fuel spent by the vessel.
We will also make two more assumptions which simplify the problem considerably: The current is always parallel to the river channel, and it only depends on the longitudinal distance from the bank -- not the lateral distance, see Fig.~\ref{fig:RiverCrossingCont}. Within this scenario, the optimal, continuous solution is the one that minimizes the total crossing time given by the functional 
\begin{equation}
\label{eq:TimeFunctional}
T[f; C] = \int_0^D \, \frac{1 + [f'(x)]^2}{\sqrt{[1+f'(x)]v^2 - [C(x)]^2} - f'(x) C(x)} \mathrm{d} x ,
\end{equation}
subject to the boundary conditions 
\begin{equation}
\label{eq:BoundaryCond}
f(0) = f(D) = 0 . 
\end{equation}
Here $f(x)$ is the path across the river, the given function $C(x)$ is the current at the distance $x$ from the initial bank and $D$ is the width of the river, i.e., the distance between the departure point and destination. 

By applying variational calculus the optimal, continuous solution may be found in a semi-analytical manner which requires surprisingly little computational effort. 
While deriving Eq.~(\ref{eq:TimeFunctional}) and implementing this optimization scheme for a specific case would be interesting enough as a student project in it is own right, the scope of our present work is \textit{quantum annealing}. For this purpose, we will \textit{discretize} the problem. The path from departure point to destination is divided into $n$ piecewise linear steps, each with equal distance $\Delta x$ in the $x$-direction. As the vessel propels from point to point it may select one out of three options: Either proceed along the straight line or to move to a distance $\Delta y$ upstream or downstream. Clearly, the latter choice is the faster one as the vessel would have to spend less time and effort to compensate for the current. However, the vessel is to arrive at a fixed destination in the end, i.e., the boundary condition Eq.~(\ref{eq:BoundaryCond}) must be fulfilled, so every deviation downstream must be compensated by a movement upstream at another point so that there is no net drift in the end. 

We may see the path as a sequence of $n$ ternary choices, e.g, $\left( \rightarrow, \nearrow, \rightarrow, \searrow, ... \right)$, which, in turn, may be listed as a ``tritstring'', i.e., a sequence of ternary numbers,
\begin{equation}
\label{eq:TritString}
\vec{s} = \left( s_1, s_2, s_3, ... \right) = \left( 0, +1, 0, -1, ... \right) .
\end{equation}
Each step is associated with a partial time-cost. By applying the midpoint rule for integration to Eq.~(\ref{eq:TimeFunctional}) for each time step and replacing $f'(x)$ in Eq.~(\ref{eq:TimeCosts}) with $\pm \Delta y/ \Delta x$ and $0$ for $s_k = \pm 1$ and $0$, respectively, we may arrive at the following time cost at step number $k$:
\begin{equation}
\label{eq:TimeCosts}
t_k(s_k) =  
\left\{
\begin{array}{ll}
\frac{\Delta x^2 + \Delta y^2}{\sqrt{(\Delta x^2 + \Delta y^2) v^2 -(\Delta x C(x_k))^2} - \Delta y C(x_k)} , & s_k=+1\\
\frac{\Delta x}{\sqrt{v^2 -[C(x_k)]^2}}, & s_k =0 \\
\frac{\Delta x^2 + \Delta y^2}{\sqrt{(\Delta x^2 + \Delta y^2) v^2 -(\Delta x C(x_k))^2} + \Delta y C(x_k)} , & s_k = -1
\end{array}
\right. ,
\end{equation}
where $x_k = (k-1/2) \Delta x$. The optimal path in this model is given by the tritstring which minimizes the total time 
\begin{equation}
\label{eq:TotalCost}
T(\vec{s}) = \sum_{k=1}^n t_k(s_k)
\end{equation}
subject to the feasibility constraint
\begin{equation}
\label{eq:BoundaryCondDiscrete}
S(\vec{s}) = \sum_{k=1}^n s_k = 0 ,
\end{equation}
which ensures no net drift. The time weights in Eq.~(\ref{eq:TimeCosts}) fulfill the inequalities 
\begin{subequations}
\label{eq:InequalitiesForTimes}
\begin{align}
\label{eq:InequalitiesForTimes_1}
& t_k(+1) > t_k(0) > t_k(-1) \\
\label{eq:InequalitiesForTimes_2}
& t_k(+1) - t_k(0) > t_k(0) - t_k(-1)
\end{align}
\end{subequations}
for every $k$. While the upper double inequality may be more or less intuitive, it may be less obvious that the lower one also holds. See App.~\ref{sec:Proof1} for a proof thereof.

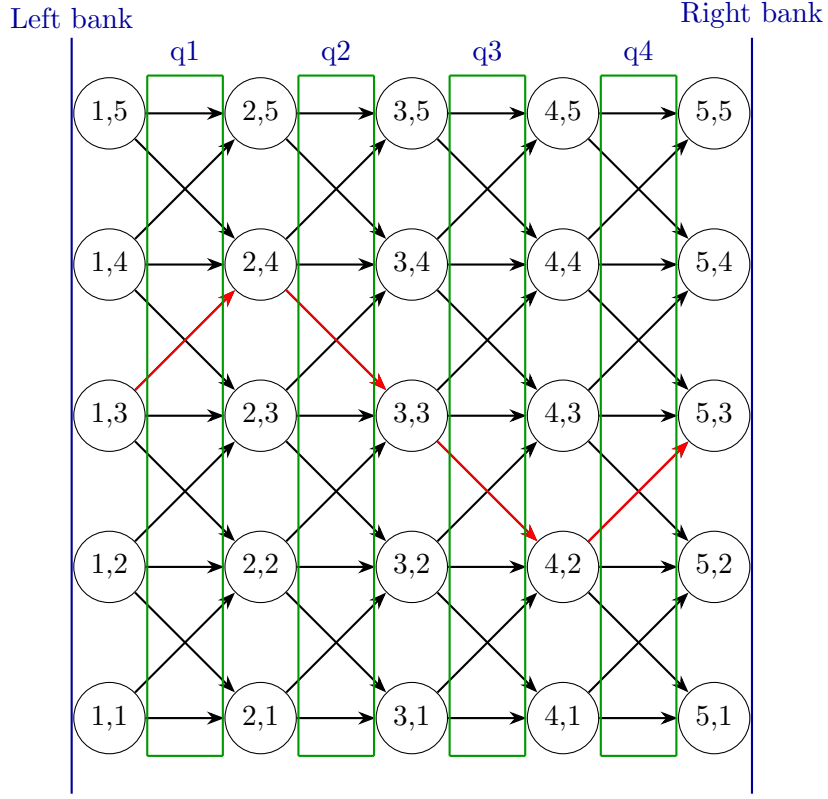
\begin{figure}
    \centering
    \begin{tikzpicture}[node distance=2cm]
        \draw[thick, blue!60!black] (1.5,1) -- (1.5,11);
        \draw[thick, blue!60!black] (10.5,1) -- (10.5,11);
        \node[blue!60!black, above] at (1.5,11) {Left bank};
        \node[blue!60!black, above] at (10.5,11) {Right bank};

        \foreach \y in {1,2,3,4,5} {
            \foreach \x in {1,2,3,4,5} {
                \node[graphnode] (n\x\y) at (2*\x,2*\y) {\x,\y};
            }
        }
        
        \foreach \x in {1,2,3,4} {
            \foreach \y in {1,2,3,4,5} {
    
                \draw[graphedge] (n\x\y) -- (n\the\numexpr\x+1\relax\y);

                \ifnum\y<5
                    \draw[graphedge] (n\x\y) -- (n\the\numexpr\x+1\relax\the\numexpr\y+1\relax);
                \fi
    
                \ifnum\y>1
                    \draw[graphedge] (n\x\y) -- (n\the\numexpr\x+1\relax\the\numexpr\y-1\relax);
                \fi
            }
        }

        \draw[graphedge, thick, red!100!black] (n13) -- (n24);
        \draw[graphedge, thick, red!100!black] (n24) -- (n33);
        \draw[graphedge, thick, red!100!black] (n33) -- (n42);
        \draw[graphedge, thick, red!100!black] (n42) -- (n53);
        \foreach \x in {1,2,3,4} {
            \node[blue!60!black, above] at (2*\x+1,10.5) {q\x};
            \draw[thick, green!60!black] (2*\x+0.5,1.5) -- (2*\x+0.5,10.5); 
            \draw[thick, green!60!black] (2*\x+0.5,1.5) -- (2*\x+1.5,1.5); 
            \draw[thick, green!60!black] (2*\x+0.5,10.5) -- (2*\x+1.5,10.5); 
            \draw[thick, green!60!black] (2*\x+1.5,1.5) -- (2*\x+1.5,10.5); 
        }
       
    \end{tikzpicture}
    \caption{In the discretized formulation, the path across the water becomes a sequence of steps upwards, forward or downwards in a structured, weighted graph. Here the path is subdivided into $n=4$ steps, the red arrows indicate one particular feasible solution: $\vec{s} = (1, -1, -1, 1)$.}
    \label{fig:graph_based_river}
\end{figure}

Our problem is a particular case of a shortest path problem for a structured, unidirectional weighted graph~\cite{Gallo1988}, see Fig.~\ref{fig:graph_based_river}.
A brute force approach to this problem would imply checking each of the $N = 3^n$ possible tritstrings. And even if we were able to single out only feasible candidates satisfying Eq.~(\ref{eq:BoundaryCondDiscrete}) somehow, this would still constitute a number of candidates which grows superpolynomially with the number of trits $n$. Thus, it may seem fair to assume that this discretized Zermelo problem also requires a time that grows exponentially in $n$. However, as we will see, it does not. 

\section{The Annealing Scheme}
\label{sec:Scheme}

Adiabatic quantum computing consists in exploiting the adiabatic theorem in quantum physics to find the solution to a optimization problem of interest. To this end, we must encode the problem into a Hamiltonian operator which has the optimal solution as the ground state. By starting out with a system in the known ground state of a different, well understood initial Hamiltonian, we will slowly change our system into the problem Hamiltonian. If this is down slowly enough, our system should remain in its ground state and our solution may be determined by reading off the resulting final state. 

With the notion of  \textit{trits} and \textit{tritstrings}, as opposed to bits and bitstrings, the natural first step in this regard is to introduce \textit{qutrits},
\begin{equation}
\label{eq:QuTrit}
|\psi \rangle = a_{+1} |+1\rangle + a_0 |0\rangle + a_{-1} | -1 \rangle, 
\end{equation}
as opposed to the more common notion of a qubit, $|\psi\rangle = a_0 |0 \rangle + a_1|1\rangle$. Any quantum system which may be described by three orthogonal, discrete quantum states may constitute a qutrit. Regardless of implementation, the system may formally be seen as that of a spin system with total spin quantum number $S=1$. While the dynamics within a spin 1/2-system may be described in terms of the Pauli matrices $\sigma_x$, $\sigma_y$ and $\sigma_z$ -- in addition to the identity matrix, the $S=1$-case may be described by the corresponding $3 \times 3$ matrices 
\begin{equation}
\label{eq:PauliMatrices}
\sigma_x = \frac{\hbar}{\sqrt{2}} 
 \left( 
\begin{array}{ccc} 0 & 1 & 0  \\ 1 & 0 & 1 \\ 0 & 1 & 0
\end{array} 
\right) ,\quad 
\sigma_y = \frac{\hbar}{\sqrt{2}} 
\left( 
\begin{array}{rrr} 0 & -i & 0  \\ i & 0 & -i \\ 0 & i & 0
\end{array} 
\right) ,\quad 
\sigma_z = 
\hbar \left( 
\begin{array}{ccc} 1 & 0 & 0  \\ 0 & 0 & 0 \\ 0 & 0 & -1
\end{array} 
\right) .
\end{equation}

Each ternary variable in the cost function, Eq.~(\ref{eq:TotalCost}), is now replaced by a qutrit, Eq.~(\ref{eq:QuTrit}). 
For a full register of $n$ qutrits, the set of all possible paths constitues the \textit{computational basis} for the $N = 3^n$-dimensional vector space in which our state $|\Psi \rangle$ resides. 

In order to encode the optimal solution into the ground state of a target Hamiltonian, we will define certain single-qutrit operators in the larger space. Specifically, we define the operator $D_k$  
as the one that acts on qutrit number $k$ with the diagonal $3 \times 3$ matrix $\mathrm{diag}(\vec{t}_k)$ where $\vec{t}_k = (t_k(+1), t_k(0), t_k(-1))$, cf. Eq.~(\ref{eq:TimeCosts}), and leaves the other $n-1$ ones untouched. It is given by a long sequence of Kronecker products where each factor is the identity matrix $I_3$ -- except for factor number $k$:
\begin{equation}
\label{eq:DefD}
D_k(\vec{t}_k)  =  I_3 \otimes ... \otimes I_3 \otimes \mathrm{diag}( \vec{t}_k ) \otimes I_3 \otimes ... \otimes I_3 =  
I_3^{\otimes(n-k)}\otimes \mathrm{diag}( \vec{t}_k ) \otimes I^{\otimes(k-1)}_3 .
\end{equation}
The large Kronecker powers of identity matrices are themselves large identity matrices, $I^{\otimes p}_3 = I_{3^p}$. While $D_k$ quickly becomes a very large matrix, it maintains a diagonal structure, which, in turn, significantly reduces the memory requirements in assigning these matrices numerically. With $D_k$ we may now encode the cost function, Eq.~(\ref{eq:TotalCost}), as the operator
\begin{equation}
\label{eq:DefHcost}
H_\mathrm{cost} = \sum_{k=1}^n D_k (\vec{t}_k) .
\end{equation}
The ground state of this diagonal Hamiltonian will correspond to the fastest path across the river when the boundary condition in Eq.~(\ref{eq:BoundaryCondDiscrete}) is not imposed. This path consistently follows the current,
$\vec{s} = \left( -1, -1, ..., -1 \right)$, which in a quantum context is written -- with brackets -- as the product state 
\begin{equation}
\label{eq:PsiInit}
|\Psi_0 \rangle = |-1, -1, ..., -1 \rangle = |-1 \rangle^{\otimes n}  .
\end{equation}

However, as we are only interested in the optimal solution subject to Eq.~(\ref{eq:BoundaryCondDiscrete}), we must construct our Hamiltonian such that the ground state abides by this. This may be achieved by augmenting our Hamiltonian with a penalty term, $H = H_\mathrm{cost} + \beta H_\mathrm{pen}$, where
\begin{equation}
\label{eq:DefHpen}
H_\mathrm{pen} = \left(\sum_{k=1}^n S_k^{(z)} \right)^2 .
\end{equation}
$S_k^{(z)}$ imposes $\sigma_z$ in Eq.~(\ref{eq:PauliMatrices}) on qutrit $k$ and leaves the rest unchanged: 
\begin{equation}
\label{eq:DefSk}
S_k^{(z)} =
I_3^{\otimes(n-k)} \otimes \sigma_z \otimes I_3^{\otimes(k-1)} .
\end{equation}
We emphasize that the square in Eq.~(\ref{eq:DefHpen}) is the ordinary matrix product, not the Kronecker product this time. Any quantum state $|\Psi\rangle$ which is an eigenstate of the total spin projection operator $\sum_{k=1}^n S_z^{(k)}$ with eigenvalue zero, is unaffected by $H_\mathrm{pen}$. Any other state, i.e., any state with a non-zero net spin projection, will, however, be affected by the penalty term. Provided that the penalty parameter $\beta$ is set sufficiently high, the ground state of the total Hamiltonian will provide the solution we are looking for -- the computational basis state which minimizes the function in Eq.~(\ref{eq:TotalCost}) subject to the constraint Eq.~(\ref{eq:BoundaryCondDiscrete}). 

The aim of our annealing process is thus to manipulate our system so that it ends up in the ground state of the final Hamiltonian $H_f = H_\mathrm{cost} + \beta H_\mathrm{pen}$. We can do this by starting out with an initial Hamiltonian, $H_i$, for which we do know the ground state and then, by exploiting the adiabatic theorem, slowly change our system from $H_i$ to $H_f$. The most common way to do so is to evolve our system according to
\begin{equation}
\label{eq:TraditionalAnneal}
H(t) = (1-s(t)) H_i + s(t) H_f ,
\end{equation}
where the time-dependent function $s(t)$, the \textit{annealing schedule}, slowly increases from $0$ to $1$~\cite{McGeoch2014}, using $H_i = -H_\mathrm{X}$ where
\begin{equation}
\label{eq:DefHexplore}
H_\mathrm{X} = \sum_{k=1}^n S_k^{(x)},
\end{equation}
with $S_k^{(x)}$ is defined analogously to $S_k^{(z)}$ in Eq.~(\ref{eq:DefSk}) with $\sigma_x$ instead of $\sigma_z$ from Eq.~(\ref{eq:PauliMatrices}). The ground state of this initial Hamiltonian is the product $|+\rangle^{\otimes n}$, the qutrit-analogue of the \textit{uniform superposition state}, where the single qutrit state $|+ \rangle = 1/2 \, \left( 1, \sqrt{2}, 1 \right)^T$ is the eigenstate of $\sigma_x$ with maximal eigenvalue.

\begin{figure}
    \centering
    \includegraphics[width=1\linewidth]{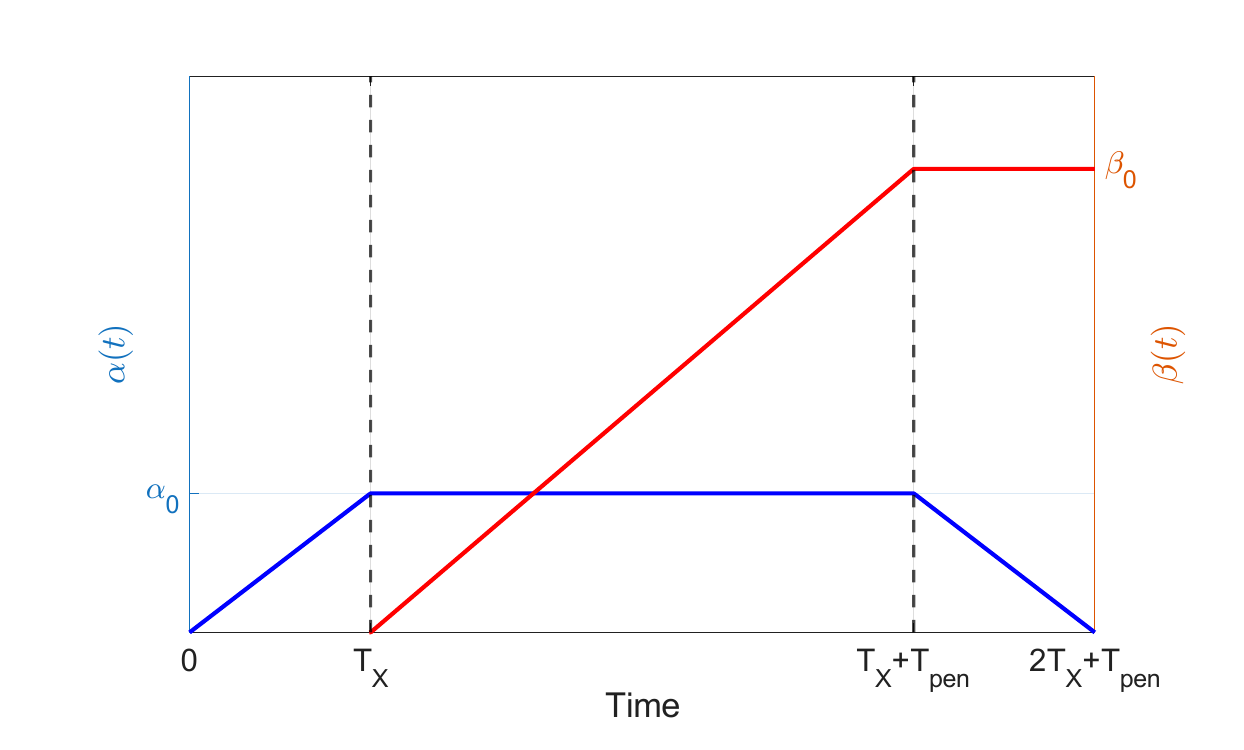}
    \caption{The annealing schedules. The anneal is divided into three phases: 1) the exploration term is ramped on, 2) the penalty is ramped on and 3) the exploration is ramped off.}
    \label{fig:Schedules}
\end{figure}

While this may be a direction, we will here apply a different annealing scheme:
\begin{equation}
\label{eq:LZannealing}
H(t) = H_\mathrm{cost} + \alpha(t) H_\mathrm{X} + \beta(t) H_\mathrm{pen} .
\end{equation}
It is implemented in three steps: Firstly, the term $H_\mathrm{X}$, which we will refer to as the \textit{exploration term}, is ramped on by increasing $\alpha(t)$ from zero to some maximum value $\alpha_\mathrm{max}$. Next, the penalty term, $H_\mathrm{pen}$ is turned on by increasing $\beta$ from zero to some adequate maximum value $\beta_\mathrm{max}$, and, finally, $\alpha(t)$ is ramped off again. See Fig.~\ref{fig:Schedules}. While this scheme may appear to be an unnecessarily cumbersome one, it does come with a clear advantage: With  piecewise linear functions $\alpha$ and $\beta$ it allows us to approximate the evolution of the system by means of the multistate Landau-Zener approximation. This, in turn, enables us to estimate the fidelity of the anneal without resorting to a full numerical solution of the equation that governs the evolution, i.e., the time time-dependent Schrödinger equation:
\begin{equation}
\label{eq:TDSE}
i \hbar \frac{\mathrm{d}}{\mathrm{d}t} |\Psi(t)\rangle = H(t) |\Psi(t) \rangle .
\end{equation}
We do, however, aim to compare the Landau-Zener prediction with the actual solution in order to validate the former -- and to check that, for sufficiently long annealing times, the remaining state is, in fact, the ground state of $H_f=H_\mathrm{cost} + \beta_\mathrm{max} H_\mathrm{pen}$. As the initial state we will take $|\Psi_0\rangle$ in Eq.~(\ref{eq:PsiInit}), the ground state of $H_\mathrm{cost}$, which in the computational basis simply reads $\left(0, \cdots, 0, 1 \right)^T$. Due to the fact that the dimension of our problem $N$ grows exponentially with the number of qutrits $n$ -- it is subject to the so-called \textit{curse of dimensionality} -- we  can only solve Eq.~(\ref{eq:TDSE}) for comparatively modest values of $n$ -- as opposed to the Landau-Zener prediction, which is easily determined with virtually no computational effort at all. 

\section{Avoided crossings and the Landau-Zener model}
\label{sec:LZpart}

\begin{figure}
    \centering
    \includegraphics[width=0.95\linewidth]{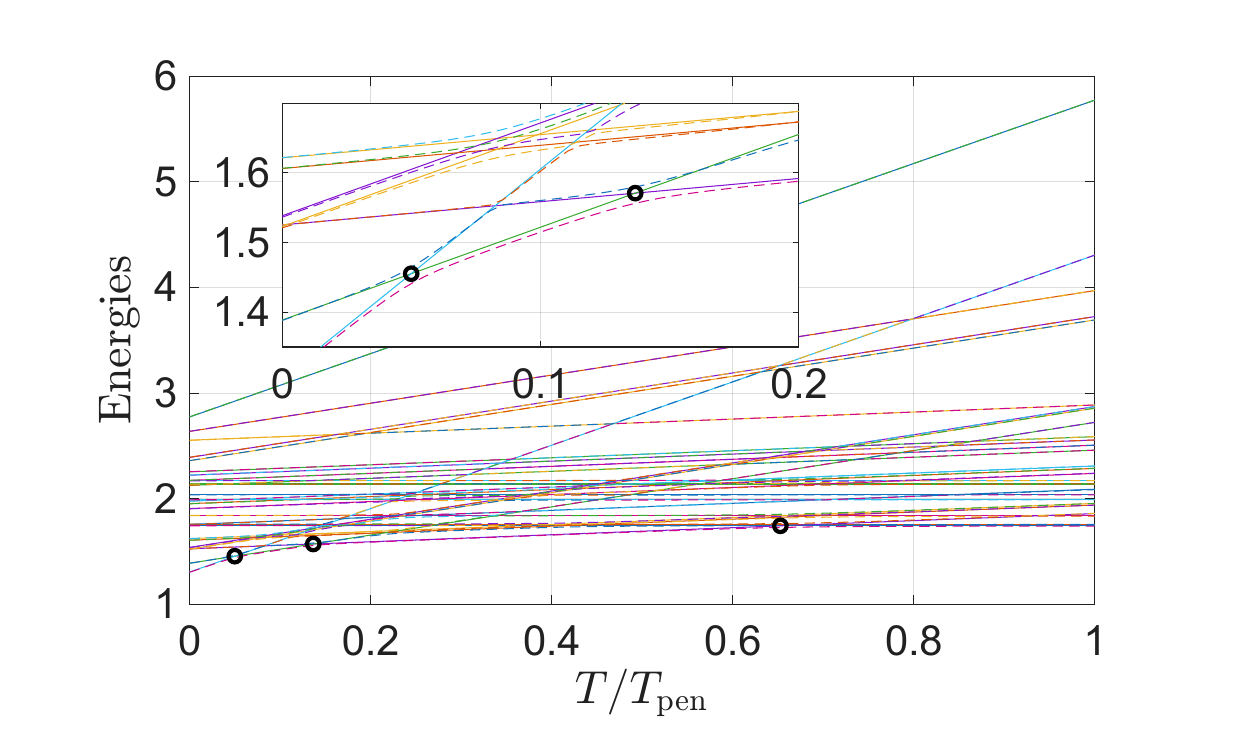}
    \caption{The diabatic (full lines) and adiabatic (dashed cures) energies as functions of the penalty strength. The energy of the feasible solutions are independent of the penalty. This case, which corresponds to $n=3$ qutrits, $\alpha = 0.05/n$ and $\beta_\mathrm{max} = 1/n$ in Eq.~(\ref{eq:LZannealing}), has three crossings between diabatic energy levels. These are indicated by black circles. The insert shows a close up on the first two (avoided) crossings.}
    \label{fig:Crossings}
\end{figure}

As the penalty term is gradually increased, the diagonal energies in the computational basis, the \textit{diabatic} basis, undergo a series of crossings. In the \textit{adiabatic} basis, i.e., the basis consisting of the instantaneous eigenstates of the time-dependent Hamiltonian $H(t)$, these crossings are generally manifested as \textit{avoided} crossings, the time-dependent eigenenergies do not cross -- provided that there is a non-zero matrix element coupling the two diabatic/computational basis states involved in the crossing. This is why we need the exploration term, 
$\alpha(t) H_\mathrm{X}$, in our dynamical Hamiltonian, Eq.~(\ref{eq:LZannealing}). Without it, we would not see any transition between crossing states, and the system would simply remain in the initial state, Eq.~(\ref{eq:PsiInit}), as the penalty is increased. With a finite coupling element between the diabatic states involved in a crossing, however, the system may -- at a certain probability -- remain in the state with the lowermost energy. In Fig.~\ref{fig:Crossings} we have displayed an example of such a scenario of (avoided) crossings with $n=3$ qutrits. The slower the (avoided) crossing is transversed and the stronger the coupling, the more likely is the system to remaining in the (adiabatic) state with lower energy. Specifically, for an isolated two-state system the probability to remain in the state with the lowermost energy after a crossing may be approximated by
\begin{equation}
\label{eq:LZprob}
P_\mathrm{LZ} = 1 - \exp\left( -\frac{2 \pi |a|^2}{\hbar b} \right) ,
\end{equation}
where $|a|$ is the coupling strength between the two diabatic states and $b$ is the slope of the difference in diagonal energies. This result is the probability of ending up in the second basis state, $(0, 1)^T$ at time $t_f$ with the model Hamiltonian
\begin{equation}
\label{eq:LZmodelHam}
H = \left(\begin{array}{cc}\frac{b}{2} t & a \\ a^{\,*} & -\frac{b}{2} t \end{array} \right) 
\end{equation}
with the initial condition $|\Psi(t_i) \rangle = (1, 0)^T$. The parameters $a$ and $b$ are constant. In the limits $t_f \rightarrow +\infty$ and $t_i \rightarrow -\infty$, Eq.~(\ref{eq:LZprob}) is, in fact, the exact solution of the Schrödinger equation, Eq.~(\ref{eq:TDSE})~\cite{Landau1932, Zener1932, Stuckelberg1932, Majorana1932}. Note that Eq.~(\ref{eq:LZprob}) is consistent with the adiabatic theorem since $P_\mathrm{LZ} \rightarrow 1$ when $b\rightarrow 0$ and $|a|>0$.

If we consider transition dynamics on finite time intervals, Eq.~(\ref{eq:LZprob}) is no longer exact. Nor is it exact when the system as a whole involves more than just two states. However, under the assumption that transitions may be approximated as local in time for two states undergoing an (avoided) crossing, it may be adapted to the multi-state case as a reasonable approximation -- provided that all relevant crossings are well separated in time. We will return to this in the following section.

Although a stronger coupling element results in a larger probability for following the lower energy path, which, in turn, would permit shorter annealing times, we will still use comparatively low values for $\alpha_\mathrm{max}$. We do so for two reasons: 1) to ensure that each adiabatic state is always dominated by a single computational basis state between crossings and, 2) to ensure that the population dynamics induced by each crossing remains well separated in time.

Now, although $H_\mathrm{X}$, contrary to $H_\mathrm{cost}$ and $H_\mathrm{pen}$, is not diagonal in the computational basis, it is still quite sparse. As seen in Fig.~\ref{fig:Crossings}, the annealing process involves a number of crossings. Thus, it may seem naïve to believe that each and every crossing is actually avoided. However, as it turns out, all crossings relevant to us, i.e., crossings in which the ground state changes from a state dominated by a certain computational basis state to another one with a lower penalty, are actually crossings for which $H_\mathrm{X}$ has a non-zero element.

In this regard we point out that since $H_\mathrm{X}$ is a single qutrit operator, any coupling element between two computational basis states which differ for more than one trit vanishes,
\begin{equation}
\label{eq:DifferTwo}
\langle s_1', ..., s_n' | H_\mathrm{X} | s_1, ..., s_n \rangle = 0
\end{equation}
whenever $s_k \neq s_k'$ for more than one $k$. Moreover, for the coupling elements to be non-zero, the qutrits in the one position which differs, must differ by one, i.e.,
\begin{equation}
\label{eq:DifferOne}
\langle s_1, ..., s_k', ..., s_n | H_\mathrm{X} | s_1, ..., s_k, ..., s_n \rangle \neq 0
\end{equation}
iff $|s_k'-s_k| = 1$. In other words, whenever the two lowermost diabatic states involved in a crossing changes one trit from $-1$ to $0$ or from $0$ to $+1$, the crossing is avoided, and the Landau-Zener formula may provide a good approximation. Now, it would hardly seem obvious that this is actually fulfilled within our model. However, Ineqs.~(\ref{eq:InequalitiesForTimes}) ensure it. A proof of this is outlined in App.~\ref{sec:Proof2}.


\section{Simulating and Adapting the Landau-Zener Formula}
\label{sec:Simulating}

In order to validate our approach, we aim to solve the time-dependent Schrödinger equation, Eq.~(\ref{eq:TDSE}), for our annealing scheme, Eq.~(\ref{eq:LZannealing}). 
There are, of course, several ways of doing so numerically. One way would be to approximate the second order Magnus operator,
\begin{equation}
\label{eq:Propagator}
|\Psi(r + \Delta t) \rangle = U(t+\Delta t, t) |\Psi(t) \rangle = e^{-i H(t+\Delta t/2) \Delta t/\hbar} |\Psi(t) \rangle + \mathcal{O}(\Delta t^3) ,
\end{equation}
by the split operator method. In this approach, the propagator may conveniently be split according to what changes in time and what is constant. In our case it natural to adapt the splitting scheme to each of the three phases involved; with the generic expressions
\begin{equation}
\label{eq:SplitOpGeneric}
U(t+\Delta t) = 
e^{-i H_A \Delta t/(2 \hbar)}  
e^{-i H_B \Delta t/\hbar}  
e^{-i H_A \Delta t/(2 \hbar)}  
+ \mathcal{O}(\Delta t^3)
\end{equation}
we have
\begin{subequations}
\label{eq:SplitOperators}
\begin{align}
\label{eq:SplitOperators1}
& H_A = H_\mathrm{cost}, \; \quad B = \alpha(t + \Delta t/2) H_\mathrm{X} \quad  \text{when} \quad  0 < t \leq T_\mathrm{X}, \\ \label{eq:SplitOperators2}
& H_A = H_\mathrm{cost} + \alpha_\mathrm{max} H_\mathrm{X}, \;\quad B = \beta(t + \Delta t/2) H_\mathrm{pen} \\ 
\nonumber
& \text{when} \quad T_\mathrm{X} < t \leq T_\mathrm{X} + T_\mathrm{pen} \quad \text{and} \\
\label{eq:SplitOperators3}
& H_A = H_\mathrm{cost} + \beta_\mathrm{max} H_\mathrm{pen}, \; \quad B = \alpha(t + \Delta t/2) H_\mathrm{X} \\
\nonumber
& 
\text{when} \quad 
T_\mathrm{X} + T_\mathrm{pen} < t \leq 2 T_\mathrm{X} + T_\mathrm{pen} .
\end{align}
\end{subequations}
Since both schedules $\alpha(t)$ and $\beta(t)$ are linear within their respective time intervals, the dynamical part of the propagator can be updated iteratively by matrix multiplications. The ramp on/off phase of the exploration term, Eq.~(\ref{eq:DefHexplore}), is computationally more costly in this regard as it involves matrix-matrix multiplications with dense matrices at each time step. The cost and penalty terms in Eqs.~(\ref{eq:DefHcost}) and (\ref{eq:DefHpen}), on the other hand, are easily accounted for as they are diagonal.

As discussed, the Landau-Zener model may provide good approximations for the relevant transition probabilities when the avoided crossings are well separated in time -- and the coupling elements are sufficiently low. 
However, in a situation which allows for multiple energy paths, the final state probability is subject to interference effects. Thus, it is crucial to keep track of the phase of each amplitude, not only its absolute value~\cite{Shevchenko2010}. This phase involves both the dynamical phase associated with the energy evolution and an ``instantaneous kick'' picked up at each avoided crossing. This latter phase shift is only well-defined in the adiabatic basis. Whether the sequence of (avoided) crossings displayed in Fig.~\ref{fig:Crossings} may be considered ``well separated'' does not only depend on the total annealing time $T_\mathrm{pen}$, but also the magnitude of the coupling $\alpha_\mathrm{max}$. As $T_\mathbf{pen}$ increases towards larger values, the notion of well-separated (avoided) crossings is increasingly justified.

In this context, we will, quite conveniently and pragmatically, circumvent the whole issue of possible interference effects. In the limit that  $T_\mathrm{pen}$ becomes large -- when the evolution becomes increasingly adiabatic, the state $|\Psi(t)\rangle$ will to an increasing extent follow the instantaneous ground state, i.e., the adiabatic state with lowest energy, at all times. As the dynamical population of any other adiabatic state vanishes, so does any interference effect. Correspondingly, if we require a high final ground state probability, which we will refer to as the \textit{fidelity} of the anneal, we may estimate it as the simple product of the probabilities to remain in the ground state for each avoided crossing, Eq.~(\ref{eq:LZprob}):
\begin{equation}
\label{eq:FidelityLZ}
F_\mathrm{LZ} = \prod_{k=1}^n P_k \quad \text{where} \quad
P_k = 1-\exp\left( -\frac{\pi \alpha_\mathrm{max}^2 T_\mathrm{pen}}{\hbar \beta_\mathrm{max}(2(n-k) + 1)}\right) .
\end{equation}
Here we have used that the coupling at each crossing is $\alpha_\mathrm{max}/\sqrt{2}$ and the slope of the diabatic energy difference for the $k$-th crossing is $\beta_\mathrm{max}/T_\mathrm{pen} \, ((n-k+1)^2 - (n-k)^2)$, cf. Eq.~(\ref{eq:LZprob}).
This estimate is to be compared with the actual fidelity,
\begin{equation}
\label{eq:Fidelity}
F = \left| \langle \vec{s}_\mathrm{opt} | \Psi(T_f) \rangle \right|^2 ,
\end{equation}
where $|\vec{s}_\mathrm{opt} \rangle$ is the computational basis state corresponding to the optimal solution and $|\Psi(T_f) \rangle$ is the final state remaining after the annealing process, which we determine by solving the Schrödinger equation~(\ref{eq:TDSE}).

\section{Results}
\label{sec:Results}

We have chosen this asymmetric Gaussian model for the current in our numerical studies:
\begin{equation}
\label{eq:RiverModel}
C(x) = C_0 \, e^{-(x-D/\pi)^2} ,
\end{equation}
where the maximum current $C_0 = 0.9 \, v$, where $v$ is the velocity of the vessel. This function determines, in turn, the costs associated with each choice of direction, cf. Eq.~(\ref{eq:TimeCosts}). 

In our numerical experiments we will conveniently take the parameters $v$ and $D$ as the units for velocity and distance, respectively. In the same spirit we set $\hbar = 1$. The stepsize in the transverse direction, $\Delta y$, is set to $0.25 \Delta x$. When it comes to the time parameters $\alpha_\mathrm{max}$ and $\beta_\mathrm{max}$ we will scale them with the number of qutrits $n$. Specifically, we have set $\alpha_\mathrm{max} = 0.05/n$ and $\beta_\mathrm{max} = 1/n$.
Scaling the penalty factor $\beta$ is motivated by Eq.~(\ref{eq:TimeCosts}); a higher density of grid points leads to smaller differences between the time weights. When also the magnitude of the exploration term is scaled with $1/n$, this too accounts, to some extent, for the fact that crossings become more numerous as $n$ increases.

\begin{figure}
\centering
\begin{tabular}{lll}
a) $T_\mathrm{pen}=100$ & b) $T_\mathrm{pen}=500$ & c) $T_\mathrm{pen}=1000$\\
\includegraphics[width=0.3\textwidth]
{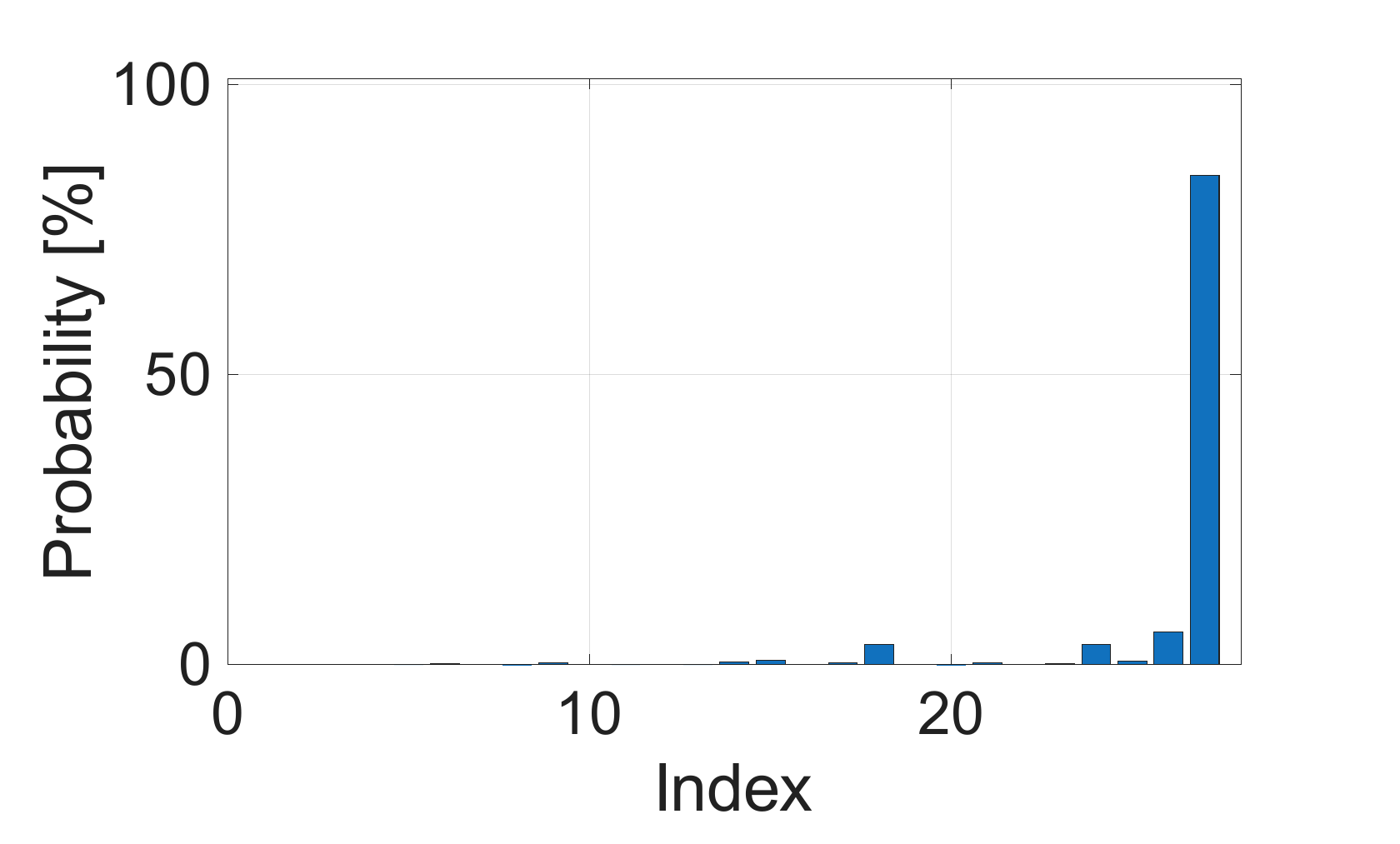} &
\includegraphics[width=0.3\textwidth]
{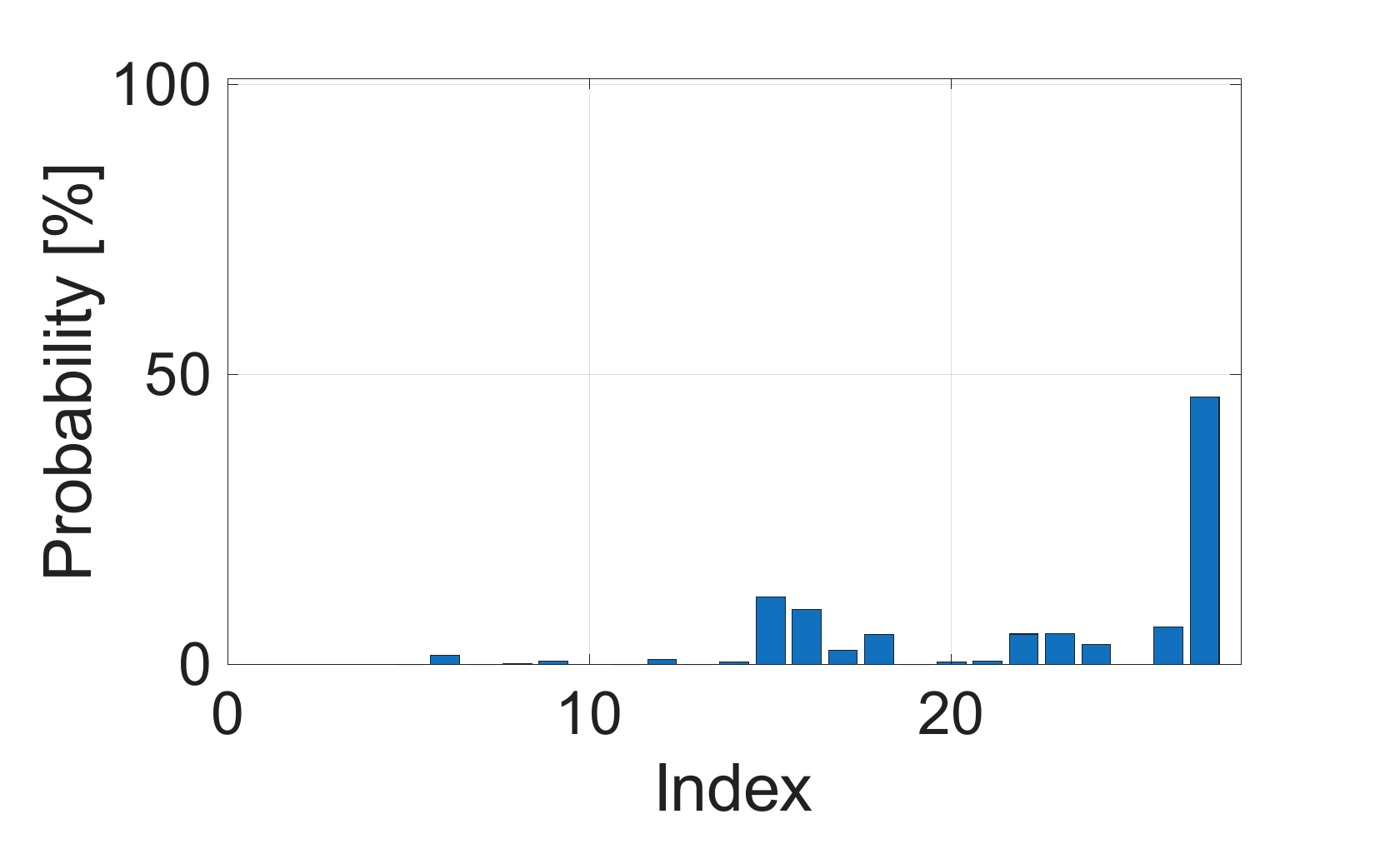}&
\includegraphics[width=0.3\textwidth]{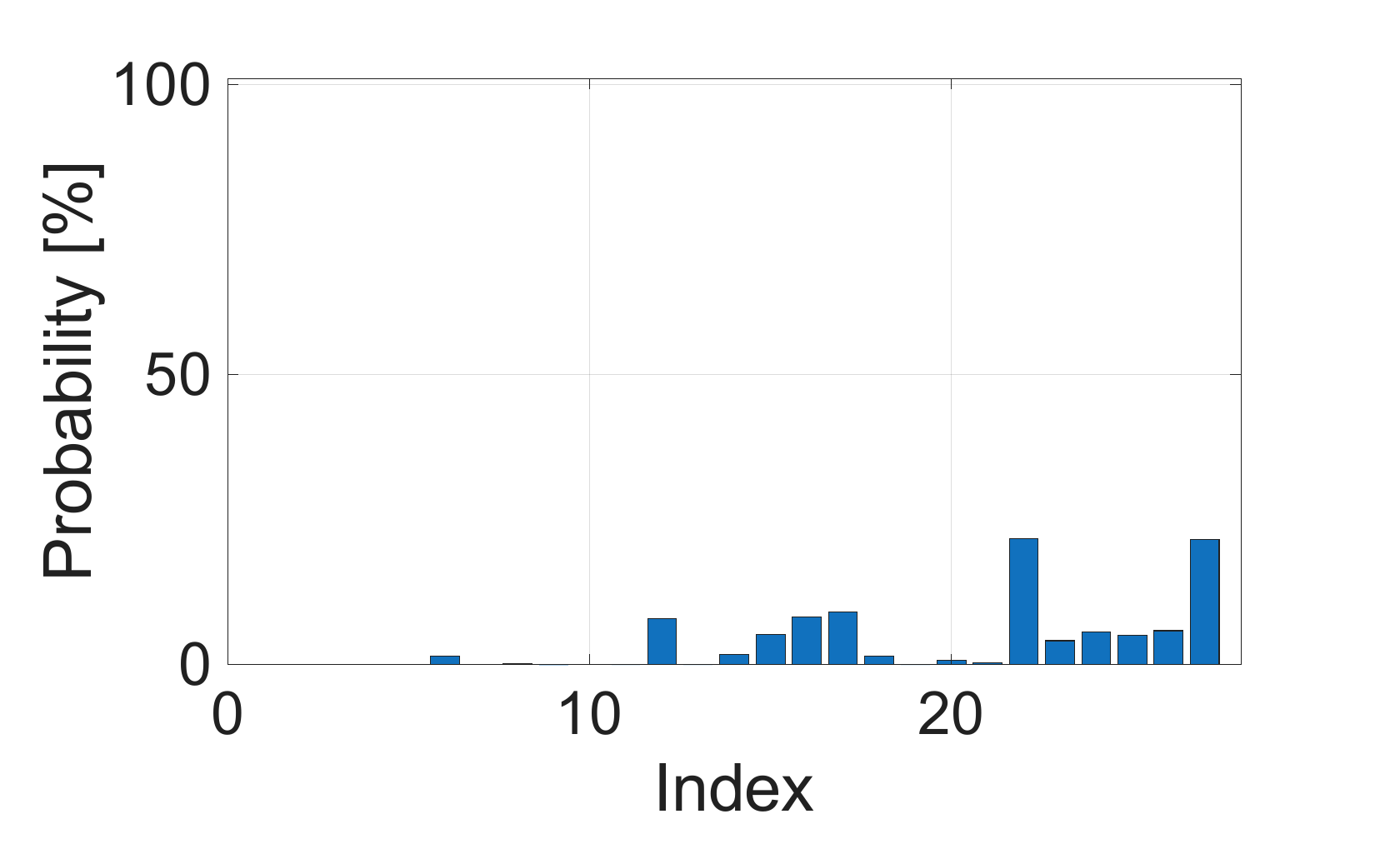} \\
d) $T_\mathrm{pen}=2000$& e) $T_\mathrm{pen}=5000$ & f) $T_\mathrm{pen}=10000$\\
\includegraphics[width=0.3\textwidth]
{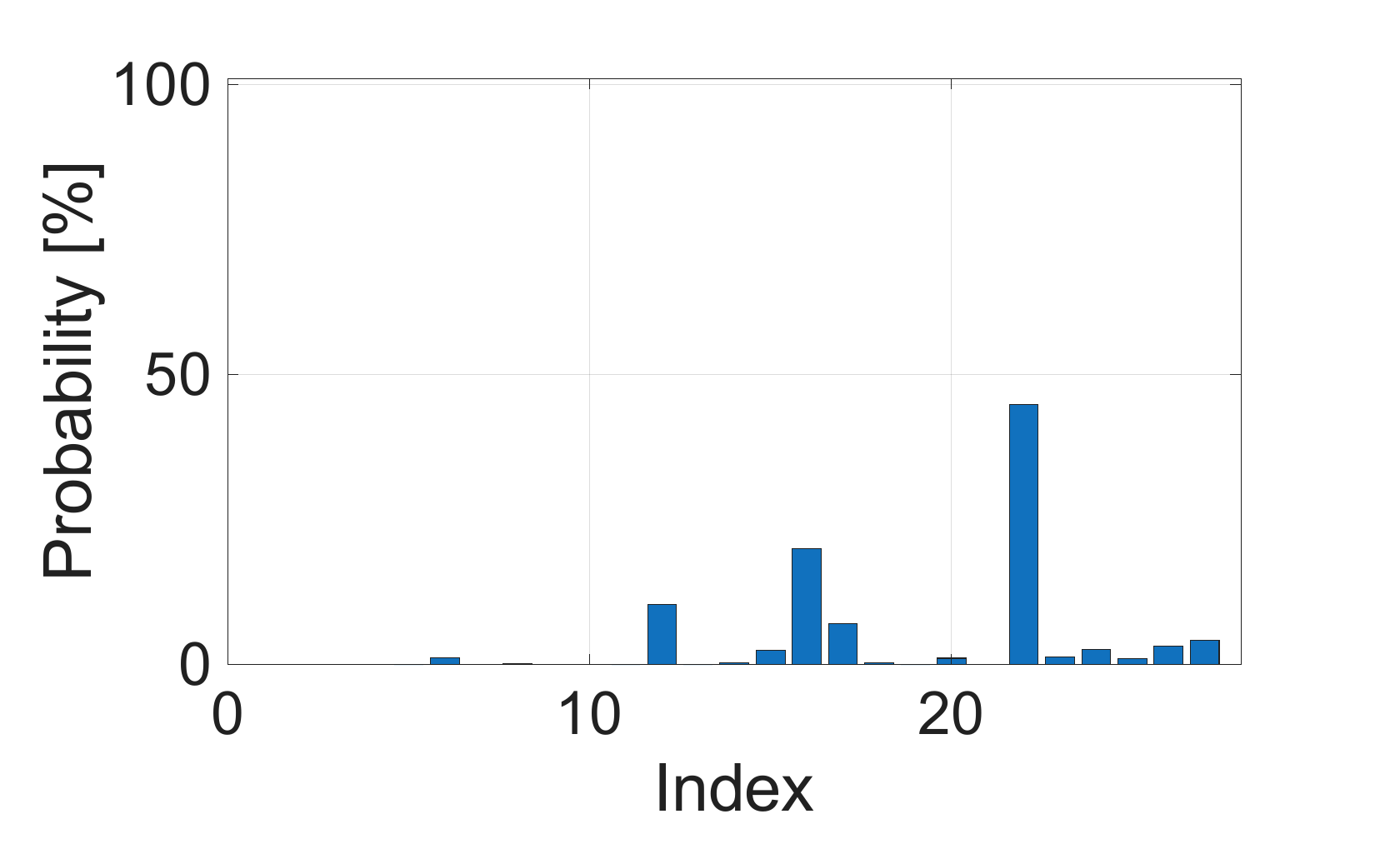}&
\includegraphics[width=0.3\textwidth]
{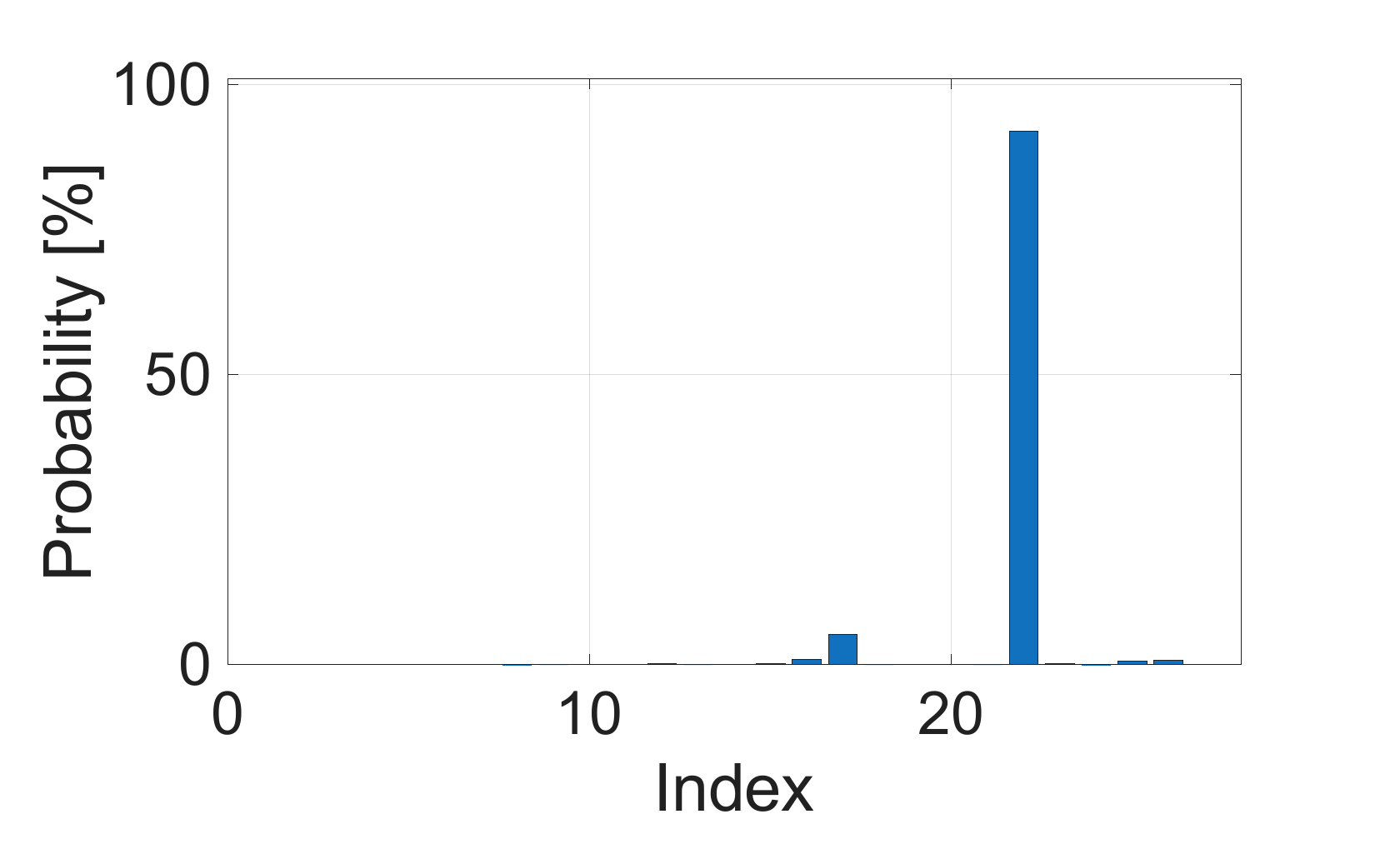}&
\includegraphics[width=0.3\textwidth]
{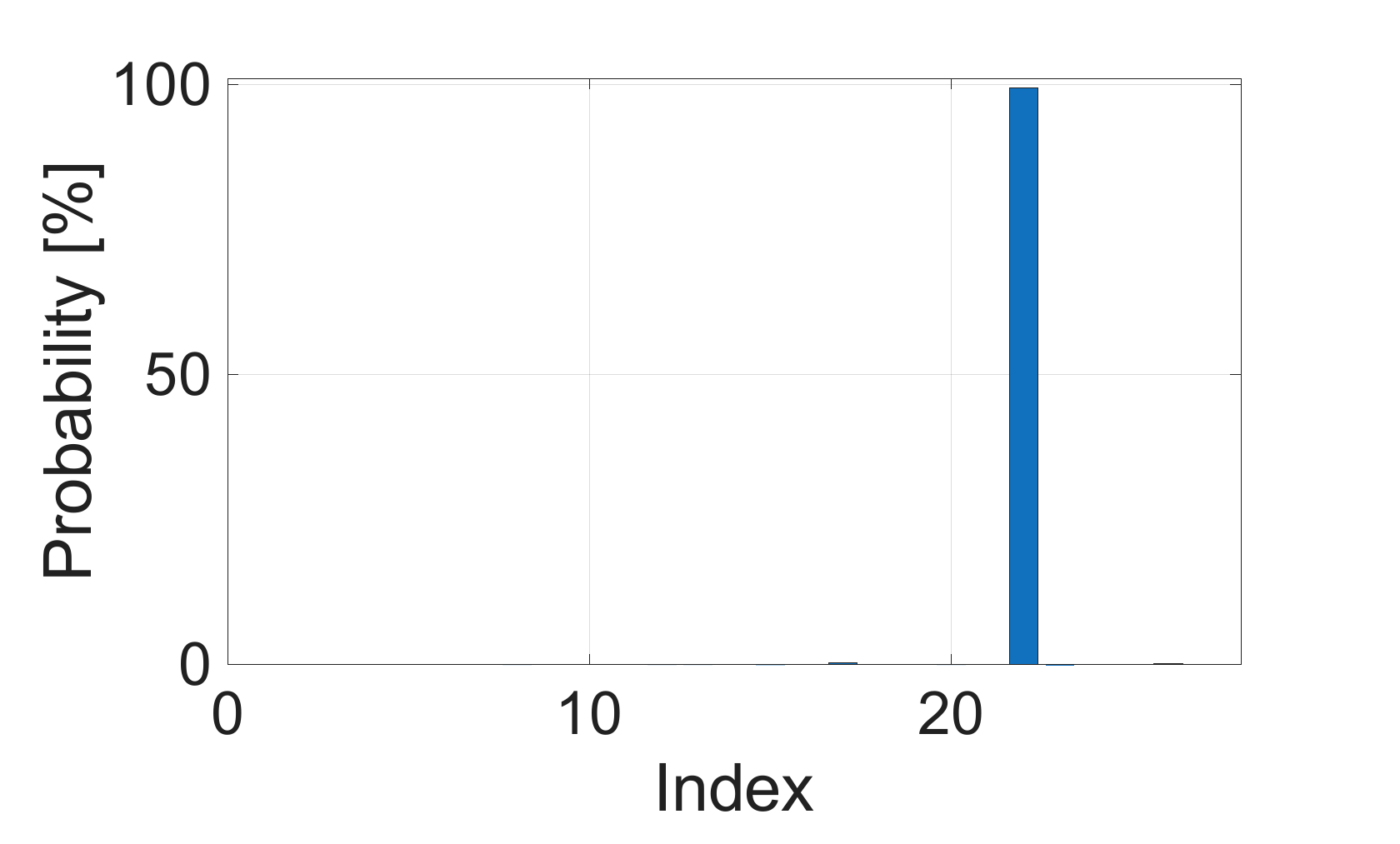} 
\end{tabular}
\caption{The probability for the various states in the computational basis for various annealing times $T_\mathrm{pen}$. The annealing times correspond, from a) to f) to $T_\mathrm{pen}=100, 500, 1000, 2000, 5000$ and $10 000$ time units, respectively. 
}
\label{fig:FinalDists}
\end{figure}

In Fig.~\ref{fig:FinalDists} we display the probability to end up in each of the computational basis states for $n=3$ qutrits for various choices of the annealing time. 
We have set $T_\mathrm{X} = T_\mathrm{pen}/5$, which ensures adiabatic evolution of the first and last phase for virtually all the times considered here.
We see that as $T_\mathrm{pen}$ increases, the fidelity increases -- in accordance with the adiabatic theorem. In this particular case, the optimal solution corresponds to state number $22$, which, in turn, is the path $\vec{s}_0= (-1, 0, 1)$, i.e., down-straight-up.

\begin{figure}
\centering
\begin{tabular}{lr}
\includegraphics[width=0.5\textwidth]{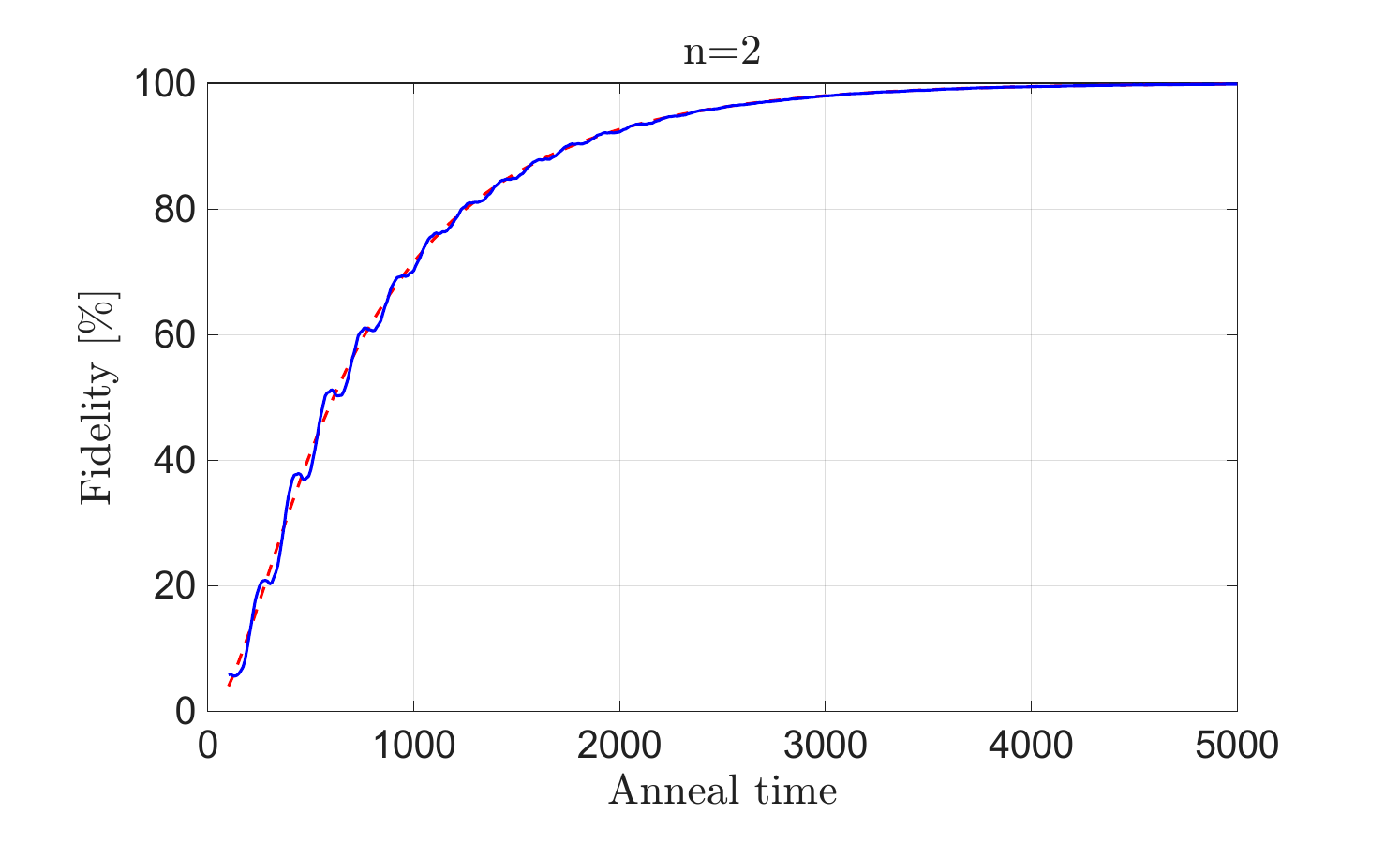} &
\includegraphics[width=0.5\textwidth]{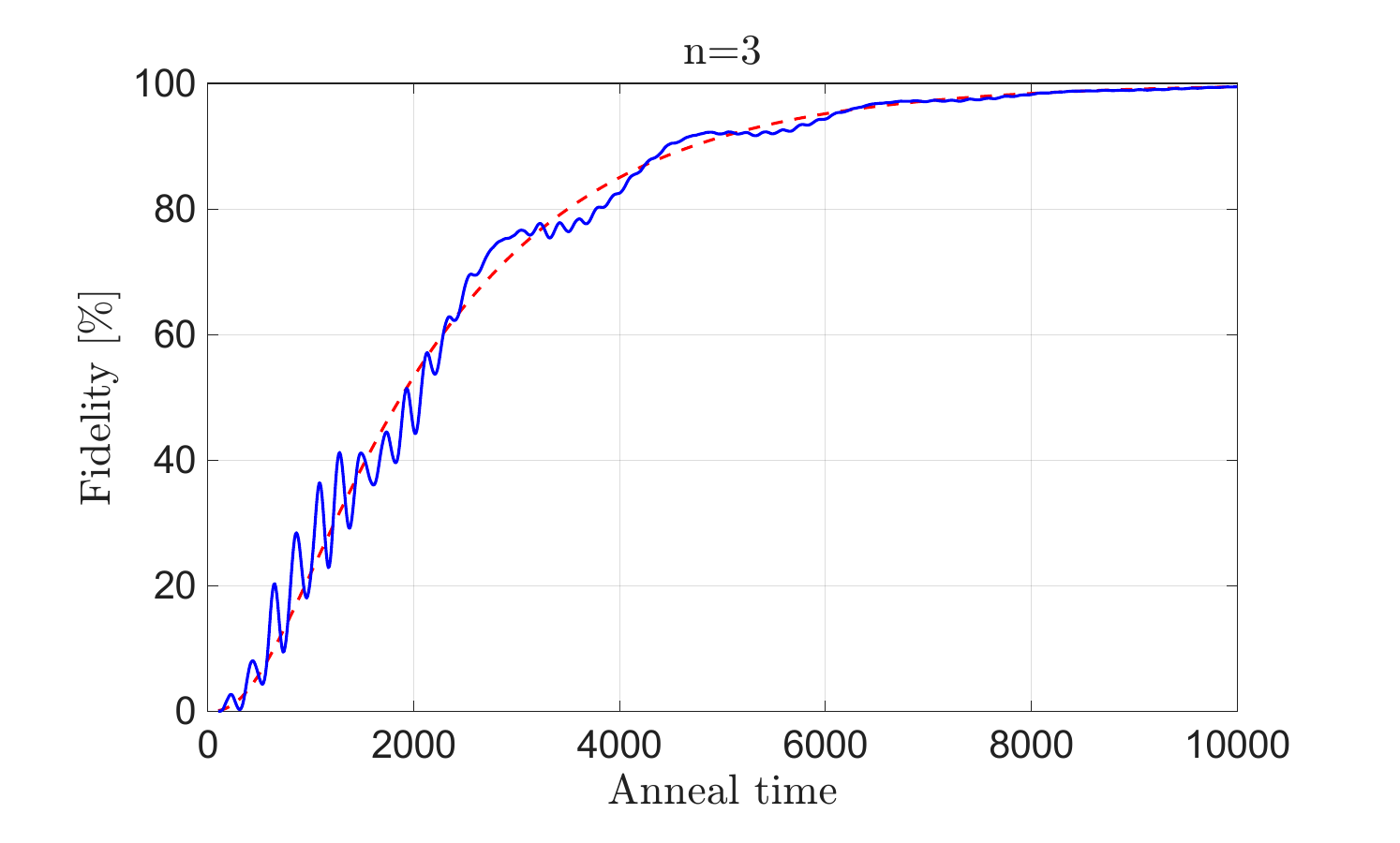} \\
\includegraphics[width=0.5\textwidth]{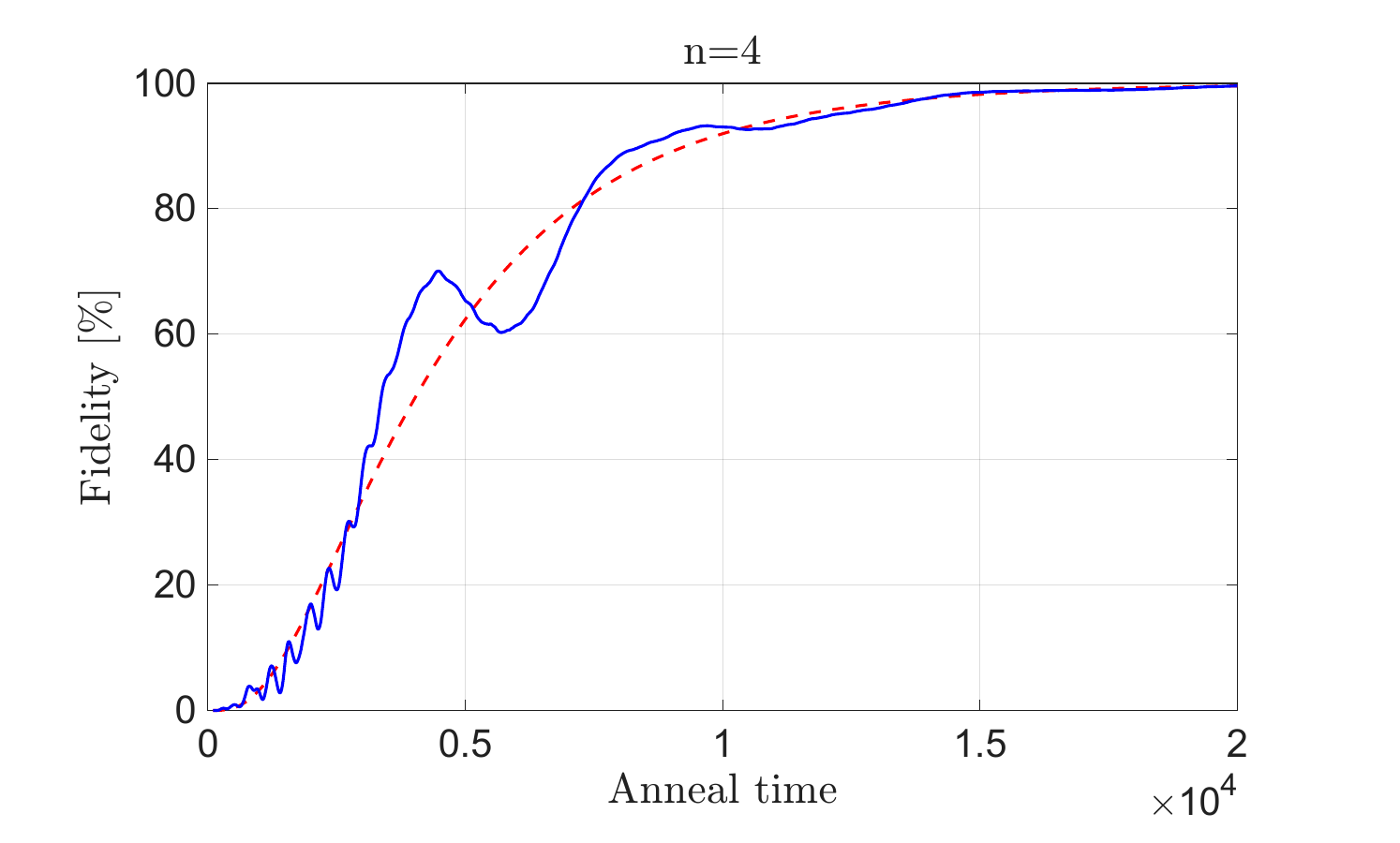} &
\includegraphics[width=0.5\textwidth]{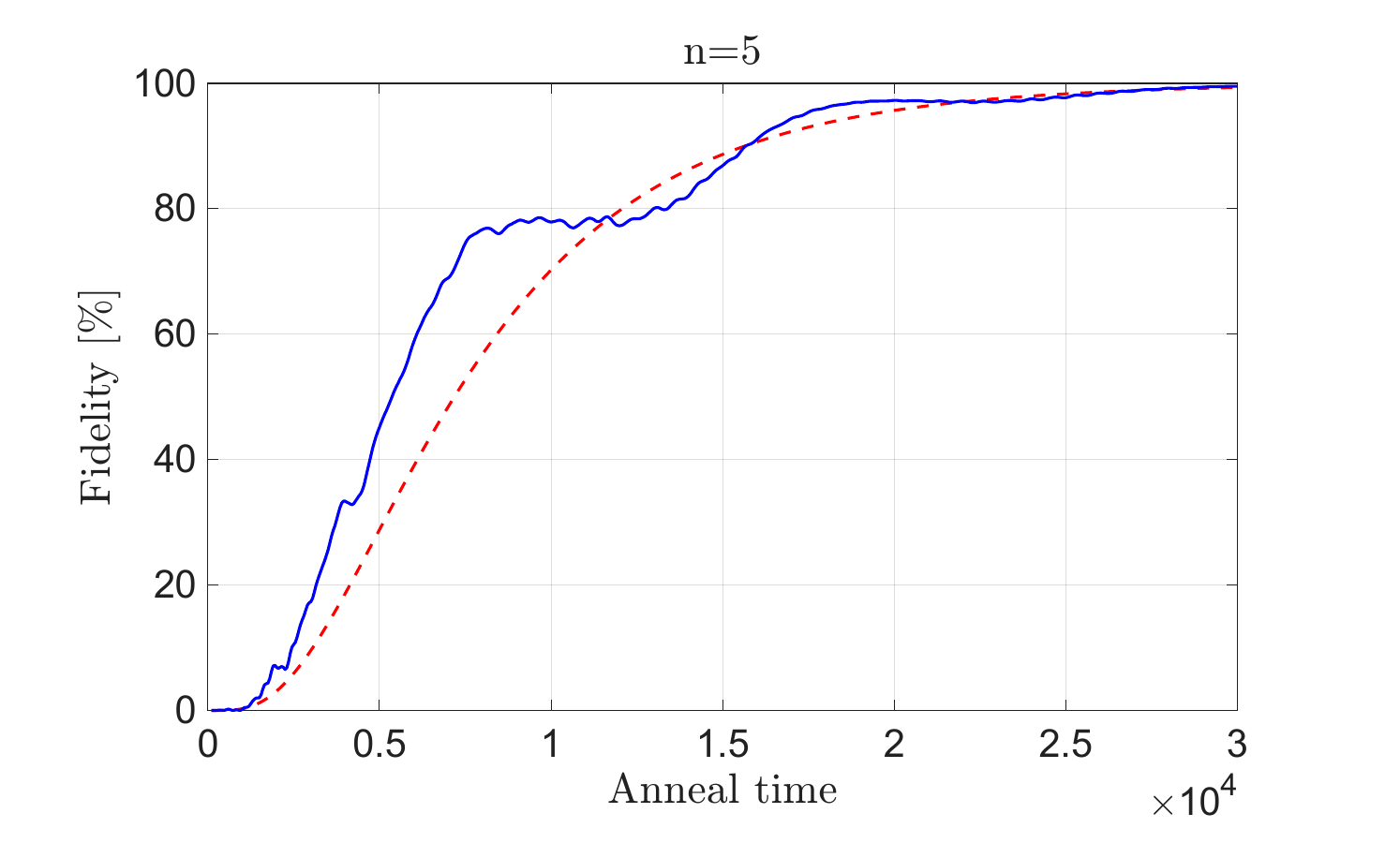}
\end{tabular}
\caption{The fidelity of the solution produced from the annealing process as a function of the penalty time $T_\mathrm{pen}$. The dashed curve is the Landau-Zener estimate, Eq.~\ref{eq:FidelityLZ}.
The full curve is the true fidelity. From top to bottom, left to right, to $n=2,3, 4$ and $5$ qutrits, respectively.  Note that the extension of the $x$-axis differ considerably in these panels.}
\label{fig:FidelityVsAnnealTime}
\end{figure}

Figure~\ref{fig:FidelityVsAnnealTime} illustrates the fidelity as a function of $T_\mathrm{pen}$ for various choices of $n$. It displays both the actual fidelity as obtained by solving the time-dependent Schrödinger equation, Eq.~(\ref{eq:TDSE}), and the corresponding Landau-Zener estimate with interference effects disregarded, Eq.~(\ref{eq:FidelityLZ}). We see that while they may deviate a bit at moderate annealing times, the actual fidelity seems to fluctuate around the Landau-Zener prediction. As the process comes close to the adiabatic limit and interferences effects become less prominent, the agreement becomes quite good, suggesting that we may very well use Eq.~(\ref{eq:FidelityLZ}) to estimate the required annealing time -- provided that we require sufficiently high fidelity.

\begin{figure}
    \centering
    \includegraphics[width=.75\linewidth]{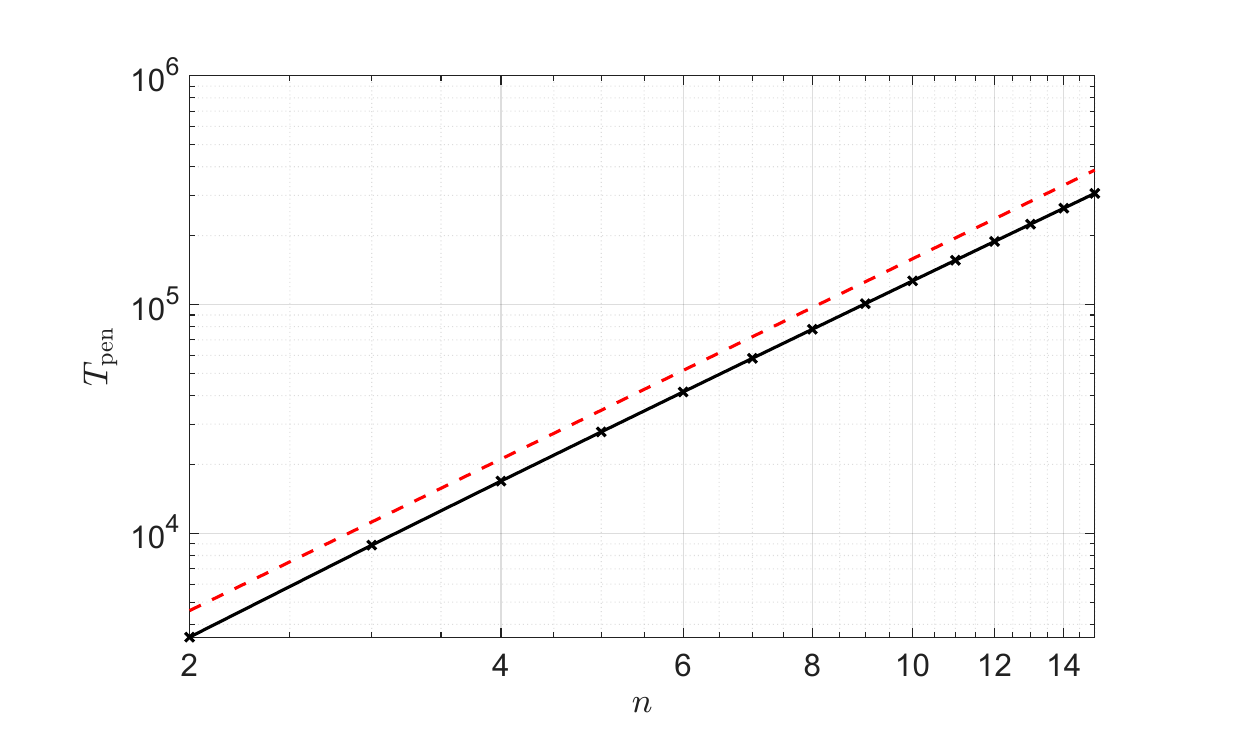}
    \caption{The time required for the the penalty onramp, $T_\mathrm{pen}$, to reach a fidelity of $99$~\% for the optimal solution. The red, dashed curve corresponds to $n^{2.2}$.}
    \label{fig:AnnealTime}
\end{figure}

In Fig.~\ref{fig:AnnealTime} we demonstrate the time it takes to reach the estimated fidely $F_\mathrm{LZ} = 99$~\% as function of $n$. We do so using logarithmic axes. 
The results suggest that the required annealing time $T_\mathrm{pen}$ follows a simple power law, $T_\mathrm{pen} \sim n^p$. The red dashed curve suggests that $p=2.2$ is a good fit.
Such a scaling is very good news for a problem assumed to be exponentially hard. 
However, the very fact that the Landau-Zener model actually works in the limit $T_\mathrm{pen} \rightarrow \infty$ also proves that this is not the case! Not only does the Landau-Zener model provide a convenient way to estimate the fidelity, it also provides a scheme for finding the optimal solution -- a \textit{classical} one: 

We take our initial tritstring $\vec{s}_{m=0}$ to be $(-1, -1, ..., -1)$, which has $S(\vec{s}_m) = -n$, cf. Eq~(\ref{eq:BoundaryCondDiscrete}). Next we determine the cost for each of the $n$ strings we get when one of the trits is increased by one. We take $\vec{s}_{m=1}$ as the one that has minimal time cost, cf. Eq.~(\ref{eq:TotalCost}). This candidate will now have $S(\vec{s}_{m=1}) = -n+1$. We iterate $n$ times to arrive at $\vec{s}_\mathrm{opt} = \vec{s}_{m=n}$, the feasible solution, $S(\vec{s}_{m=n})=0$, with minimal time cost. 
As this constitutes checking up to $n$ ternary variables $n$ times, it involves $n^2$ operations. In other words: The problem may be solved classically with very little computational effort. 
The algorithm is provided in Altorithm~\ref{algo:LZalgo}, where ``$\vec{s}[k]$'' refers to entry number $k$ of $\vec{s}$, and ``$T(\vec{s})$'' refers to Eq.~(\ref{eq:TotalCost}).

\begin{algorithm}
\SetAlgoLined
\KwData{Input size $n$ and weights $t_k(s_k)$ for $k=1,\dots,n$, $s_k \in \{-1,0,1\}$}
\KwResult{$\vec{s}_n$}
 
$\vec{s}_0 \gets (-1, \dots, -1)$ \tcp*[l]{Initialization}
\For{$m \gets 1$ \KwTo $n$}{
    \For{$k \gets 1$ \KwTo $n$}{
        \If{$\vec{s}_{m-1}[k] < 1$}{
            $\vec{s}^{\,*} \gets \vec{s}_{m-1}$ \;
            $\vec{s}^{\,*}[k] \gets \vec{s}_{m-1}[k]+1$\; 

            \uIf{$k = 1$}{
                $T_\mathrm{min} \gets T(\vec{s}^{\,*})$ \;
            }\uElseIf{$T(\vec{s}^{\,*}) < T_\mathrm{min}$}{
                $T_\mathrm{min} \gets T(\vec{s}^{\,*})$ \;
                $\vec{s}_m \gets \vec{s}^{\,*}$ \;
            }
        }
    }
}    
\caption{Optimization by repeatedly increasing individual trits}
\label{algo:LZalgo}
\end{algorithm}
 
While it is clear that this scheme provides \textit{a} feasible solution, it may not be equally obvious from the outset that this candidate actually is \textit{the} optimal solution. However, by making the ``detour'' of formulating our Landau-Zener inspired quantum annealing scheme and invoking the adiabatic theorem, it may appear a lot more intuitive.

In fact, within the context of \textit{dynamic programming}\cite{Bellman1965, Bertsekas2017} we may find another classical algorithm which scales equally favorably in terms of evaluations without any ``quantum detour''.
In order to minimize the cost function Eq.~(\ref{eq:TotalCost}) with the condition Eq.~(\ref{eq:BoundaryCondDiscrete}), we define the \textit{minimum partial cost} $C_{m, q}$ as the minimum value of 
\begin{equation}
\label{eq:PartialCost}
\sum_{k=1}^m t_k(s_k)
\end{equation}
over all assignments of the first $m$ trits such that 
\begin{equation}
\label{eq:PartialCostDeviation}
\sum_{k=1}^m s_k = q
\end{equation}
where $1 \leq m \leq n$ and $-m \leq q \leq m$. The solution is then the path that minimizes the total crossing time with $S=0$, i.e., the one corresponding to $C_{m=n,q=0}$. 
The precise algorithm is provided in Algorithm~\ref{algo:DPalgo}.

\begin{algorithm}
\SetAlgoLined
\KwData{Input size $n$ and weights $t_m(s_k)$ for $m=1,\dots,n$, $s_k \in \{-1,0,1\}$}
\KwResult{$\vec{s}_0$}
\tcp*[l]{Initialization}
$C_{0,0} \gets 0$ \;
$\vec{s}_0 \gets [\,]$

\For{$m \gets 1$ \KwTo $n$}{
    \For{$q \gets -m$ \KwTo $m$}{
        $s_k^\ast \gets \operatorname*{argmin}_{s_k \in \{-1,0,1\}}
        \bigl(C_{m-1,q-s_k}+t_m(s_k)\bigr)$

        $C_{m,q} \gets C_{m-1,q-s_k^\ast}+t_m(s_k^\ast)$

        $\vec{s}_q \gets \mathrm{Append}(\vec{s}_{q-s_k^\ast},\, s_k^\ast)$
    }
}

\caption{Dynamic programming with path tracking}
\label{algo:DPalgo}
\end{algorithm}

At each step we determine the path sequence which is minimal for each possible value of $q$. It does so by augmenting the path array $\vec{s}_q$ with the the ternary variable which minimizes the next partial cost for the $q$-value in question; from the candidates $(\vec{s}_{q+1}, -1)$, $(\vec{s}_q, 0)$ and $(\vec{s}_{q-1}, +1)$ it selects the tritstring that gives the lowest partial cost in Eq.~(\ref{eq:PartialCost}) at iteration number $m$. 
%
After having iterated from $m=1$ to $m=n$, the optimal feasible solution is provided by $\vec{s}_{q=0}$.
The computational complexity scales with the number of partial sums, which is $\sum_{m=1}^n \sum_{q=-m}^m 1 = \mathcal{O}(n^2)$. In terms of memory, on the other hand, this approach is somewhat more expensive than Algorithm~\ref{algo:LZalgo} as it stores $2n+1$ tritstrings of length $n$ as opposed to a single one in the latter case. 

\section{Discussion}
\label{sec:Discussion}

This discretized version of Zermelo's navigation problem offers several insights of interest. Perhaps the most prominent one is the realization that a quantum annealing formulation of the problem reveals an efficient \textit{classical} solution, one that was not obvious in the first place (at least not to us, the authors). Hopefully, the future holds more such realizations in store -- in addition to the discovery of novel quantum annealing schemes which provide genuine quantum advantages.

For this to come about, however, it is crucial that today's learners -- tomorrow's discoverers -- get the chance to familiarize themselves with the notion of quantum computing in general and quantum annealing/adiabatic quantum computing in particular. Relevant projects and practical implementations are key educational ingredients in this regard. In assigning such tasks, it may be tempting to fall back on old habits and resort to rather standard examples such as \textit{max cut}, \textit{sum partitioning} or other QUBO formulations using qubits~\cite{Glover2019}. Moreover, the initial state is typically the uniform superposition state, the ground state of the negative of the qubit-analogue of $H_\mathrm{X}$ in Eq.~(\ref{eq:DefHexplore}).

In this context, the discretized Zermelo problem offers a fresh and far less conventional approach to adiabatic quantum computing. The non-conventionality is not limited to the fact that it makes use of qutrits as opposed to qubits. It also has a different starting point -- one that makes sense from a physical point of view. The optimal route across the river is bit by bit, or, rather, trit by trit, moved towards the optimal \textit{feasible} route by adiabatically increasing the penalty, after having ramped on the exploration term $H_\mathrm{X}$. This, in turn, allows for a very structured sequence of avoided crossings, one that allows us to predict the fidelity, or conversely, the time required for given fidelity requirement, of the anneal in an analytical, convenient manner through the Landau-Zener formula.

As a bonus, the learner who implements this scheme gains familiarity with the multi-state Landau-Zener model. Another benefit would be gaining experience in solving the time-dependent Schr{\"o}dinger equation numerically -- an experience which is likely to confront the learner with the infamous curse of dimentionality, which, as mentoined, was the original motivating for building quantum computers~\cite{Feynman1982}.

To conclude, we have presented a simple, analytically tractable example of quantum annealing. This particular formulation allows us to employ the Landau-Zener model in order to predict the fidelity of the anneal and, thus, evade the need for a full solution of the Schr{\"o}dinger equation when predicting the annealing times. Although the problem may appear to be exponentially hard initially, the solution formulated in terms of adiabatic quantum computing reveals that the problem actually scales only quadratically -- classically. The problem -- and its solution in terms of simulating the quantum process, offers several advantages as possible student projects.

%

\appendix

\section{Proof of the asymmetry condition Ineq.~(\ref{eq:InequalitiesForTimes_2})}
\label{sec:Proof1}

If we, for convenience, omit the $k$-index, the inequality reads
\begin{equation}
\label{eq:BigUglyIneq}
t(+1) - t(0) > t(0) - t(-1) \Leftrightarrow t(1) - t(-1) >  2 t(0) ,
\end{equation}
which is equivalent to 
\begin{equation}
\label{eq:Proof1_2}
\frac{\Delta x^2 + \Delta y^2}{\sqrt{(\Delta x^2 + \Delta y^2) v^2 - (\Delta x C)^2} - \Delta y C} + 
\frac{\Delta x^2 + \Delta y^2}{\sqrt{(\Delta x^2 + \Delta y^2) v^2 - (\Delta x C)^2} + \Delta y C} >
2 \frac{\Delta x}{ \sqrt{v^2 - C^2}} .
\end{equation}
The left hand side may be recast into
\[
2 \frac{\sqrt{(\Delta x^2 + \Delta y^2)v^2 - (\Delta x C)^2}}{v^2 - C^2}
\]
so that the condition may be written
\begin{align}
\nonumber &
\frac{\sqrt{(\Delta x^2 + \Delta y^2) v^2 - (\Delta x C)^2}}{v^2 - C^2} > \frac{\Delta x}{\sqrt{v^2 - C^2}}
\Leftrightarrow \\ &
\label{eq:AlmostDoneIneq}
\sqrt{(\Delta x^2 + \Delta y^2) v^2 - (\Delta x C)^2} > \Delta x \sqrt{v^2 - C^2} .
\end{align}
Squaring both sides we arrive at
\begin{equation}
(\Delta x^2 + \Delta y^2) v^2 - (\Delta x C)^2 > \Delta x^2 (v^2 - C^2)
\Leftrightarrow 
\Delta y^2 v^2 > 0 , 
\end{equation}
which obviously holds.

\section{Why all ground state crossings are avoided}
\label{sec:Proof2}

In regard to Eq.~(\ref{eq:DifferTwo}) we claimed that whenever the penalty factor $\beta$ of Eq.~(\ref{eq:LZannealing}) increases such that the computational basis state corresponding to the minimal diagonal energy of $H_\mathrm{cost} + \beta H_\mathrm{pen}$ changes, i.e., when there is a crossing between minimal diabatic diagonal energies, the coupling element of the exploration term, $H_X$, is always non-zero. As discussed, these matrix elements are non-zero iff the labels of the corresponding computational basis states differ in one digit only -- by one, cf. Eqs.~(\ref{eq:DifferOne}) and (\ref{eq:DifferTwo}). Here we will explain why this is so. Since we will exclusively be referring to the diabatic/computational basis, which coincides with the adiabatic basis in the limit $\alpha \rightarrow 0$, the arguments will be entirely classical.

We set out to identify the tritstring $\vec{s}^{\,(q)}$ which minimizes the total cost $T(\vec{s})$ in Eq.(\ref{eq:TotalCost}) subject to the constraint $S(\vec{s}) = \sum_{k=1}^n s_k = q$, see Eq.~(\ref{eq:BoundaryCondDiscrete}). For the initial $\vec{s}^{\,(q=-n)}$, all ternary variables $s_k = -1$. We argue that $\vec{s}^{\,(q+1)}$ is constructed from $\vec{s}^{\,(q)}$ by increasing the ternary variable number $\kappa$ by one, where $\kappa$ corresponds to the index which induces the minimal increase in the total cost in going from $S=q$ to $S=q+1$:
\begin{equation}
\label{eq:DefKappa}
\kappa = \operatorname*{argmin}_{k, s^{(q)}_k<1} \left(t_k(s^{(q)}_k+1) - t_k(s^{(q)}_k) \right) .
\end{equation}
All other ternary variables remain the same for $\vec{s}^{\,(q)}$ and $\vec{s}^{\,(q+1)}$:
\begin{equation}
\label{eq:DifferenceQandQplus1}
s^{(q)}_k = s^{(q+1)}_k \quad \forall \quad k \neq \kappa \quad \text{and} \quad  s^{(q+1)}_\kappa = s^{(q)}_\kappa + 1 .
\end{equation}

In order to show that any other state vector $\vec{u}^{\,(q+1)}$ which also satisfies $S(\vec{u}^{\,(q+1)})=q+1$ has $T(\vec{u}^{\,(q+1)})>T(\vec{s}^{\,(q+1)})$, let us first consider a candidate state $\vec{u}^{\,(q+1)}$ which differs from $\vec{s}^{\,(q+1)}$ for two ternary variables such that
\begin{subequations}
\label{eq:DifferenceStarAndPrime}  
\begin{align}
\label{eq:DifferenceStarAndPrimeA}  
& u^{(q+1)}_k = s^{(q+1)}_k \quad \forall \quad k \neq \kappa, \lambda \\
\label{eq:DifferenceStarAndPrimeB}  
& u^{(q+1)}_\kappa = s^{(q+1)}_\kappa + 1 = s^{(q)}_\kappa + 2  \\
\label{eq:DifferenceStarAndPrimeC}  
& u^{(q+1)}_\lambda = s^{(q+1)}_\lambda - 1 = s^{(q)}_\lambda - 1 .
\end{align}
\end{subequations}
In other words, instead of a single excitation of the ternary variable $s^{(q)}_\kappa$, $\vec{u}^{\,(q+1)}$ involves a double excitation of $s^{(q)}_\kappa$ and a single de-excitation of some other ternary variable $s^{(q)}_\lambda$. In this case, Eq.~(\ref{eq:DifferenceStarAndPrimeB}) requires that $u^{(q+1)}_\kappa = +1$, $s^{(q+1)}_\kappa = 0$ and $s^{(q)}_\kappa = -1$. Eq.~(\ref{eq:DifferenceStarAndPrimeC}) requires that $s^{(q+1)}_\lambda > -1$. This implies that $s^{(q')}_\lambda$ was increased for a $q'<q$, which, in turn, means that 
\begin{equation}
\label{eq:HistoryCrit}
t_\lambda(s^{(q+1)}_\lambda) - t_\lambda(s^{(q+1)}_\lambda-1) < t_\kappa(s^{(q+1)}_\kappa) - t_\kappa(s^{(q+1)}_\kappa-1) .
\end{equation}
Otherwise, this would violate Eq.~(\ref{eq:DefKappa}) since $\lambda \neq \kappa$. In our case, since $s^{(q+1)}_\kappa = 0$, we also have $t_\kappa(s^{(q+1)}_\kappa) - t_\kappa(s^{(q+1)}_\kappa-1) =   t_\kappa(0) - t_\kappa(-1)$.
We also note that by virtue of Ineq.~(\ref{eq:InequalitiesForTimes_1}) any difference $t_k(s_k) - t_k(s_k-1)$ is strictly positive for any $k$.
The difference in time cost for the two states $\vec{u}^{\,(q+1)}$ and $\vec{s}^{\,(q+1)}$ may now be written
\begin{align}
\nonumber
& T(\vec{u}^{\,(q+1)}) - T(\vec{s}^{\,(q+1)}) = 
t_\kappa(u^{(q+1)}_\kappa) - t_\kappa(s^{(q+1)}_\kappa) + 
t_\lambda(u^{(q+1)}_\lambda) - t_\lambda(s^{(q+1)}_\lambda) =
\\ &
\nonumber
t_\kappa(s^{(q+1)}_\kappa+1) - t_\kappa(s^{(q+1)}_\kappa) + 
t_\lambda(s^{(q)}_\lambda-1) - t_\lambda(s^{(q)}_\lambda) = 
\\ &
\label{eq:DifferenceLarger}
t_\kappa(+1) - t_\kappa(0) -  
( t_\lambda(s^{(q)}_\lambda) - t_\lambda(s^{(q)}_\lambda-1)) .
\end{align}
With Ineqs.~(\ref{eq:InequalitiesForTimes_2}) and (\ref{eq:HistoryCrit}) we may see that
\begin{align}
\nonumber
& T(\vec{u}^{(\,q+1)}) - T(\vec{s}^{\,(q+1)}) >
t_\kappa(0) - t_\kappa(1) - (t_\kappa(0) - t_\kappa(-1))
 = 0  \Leftrightarrow 
 \\ & 
\label{eq:ConclusionPart1}
 T(\vec{u}^{\,(q+1)}) > T(\vec{s}^{\,(q+1)}) .
\end{align}
We emphasize that Ineq.~(\ref{eq:InequalitiesForTimes_2}) is crucial for this to hold.

Of course, other candidates $\vec{u}^{\,(q+1)}$ should be inspected as well. We may construct $\vec{u}^{\,(q+1)}$ from $\vec{s}^{\,(q)}$ by, in addition to increasing $s^{(q)}_\kappa$ by one, increasing ternary variable number $\lambda$ by one, $u^{(q+1)}_\lambda = s^{(q)}_\lambda+1$, and decreasing variable number $\mu$, $u^{(q+1)}_\mu = s^{(q)}_\mu-1$, while keeping the remaining $n-3$ variables of $\vec{s}^{\,(q)}$ unaltered. Analogously to the previous case, the decrease in $T(\vec{u}^{\,(q+1)})$ relative to $T(\vec{s}^{\,(q+1)})$ that the reduction in $s^{(q)}_\mu$ leads to, is now overcompensated by the increase in $s^{(q)}_\lambda$ -- because of Eq.~(\ref{eq:HistoryCrit}).

In this way it may also be seen that increasing more single-excitations and de-excitations in a manner which increases $q$ only makes things worse.

Thus, whenever $\beta$ reaches a crossing value, or formulated more technically: a critical value $\beta_c$ such that
\begin{equation}
\lim_{\beta \rightarrow \beta_c^-} \operatorname*{argmin}_{\vec{s} \in \{\pm 1, 0\}^n}  \left[ T(\vec{s}) + \beta (S(\vec{s}))^2 \right] \neq
\lim_{\beta \rightarrow \beta_c^+} \operatorname*{argmin}_{\vec{s} \in \{\pm 1, 0\}^n} \left[ T(\vec{s}) + \beta (S(\vec{s}))^2 \right],
\end{equation}
the two limits will differ only for one ternary variable -- by exactly one. Correspondingly, whenever the computational basis state corresponding to the ground state (in the limit $\alpha \rightarrow 0$) changes, the matrix element corresponding to the coupling element between the two temporary ground states is non-zero. This outline also shows why the classical Algorithm~\ref{algo:LZalgo} indeed produces the optimal feasible solution as $\vec{s}^{\,(q=0)}$.

\bibliographystyle{apsrev4-2} 
\bibliography{References}     

@article{Landau1932,
  author  = {Landau, Lev Davidovich},
  title   = {Zur Theorie der Energie{\"u}bertragung. II},
  journal = {Physikalische Zeitschrift der Sowjetunion},
  volume  = {2},
  pages   = {46--51},
  year    = {1932}
}

@article{Zener1932,
  author  = {Zener, Clarence},
  title   = {Non-Adiabatic Crossing of Energy Levels},
  journal = {Proceedings of the Royal Society of London. Series A},
  volume  = {137},
  pages   = {696--702},
  year    = {1932}
}

@article{Stuckelberg1932,
  author  = {E. C. G. St{\"u}ckelberg},
  title   = {Theorie der unelastischen St{\"o}{\ss}e zwischen Atomen},
  journal = {Helvetica Physica Acta},
  volume  = {5},
  pages   = {369--422},
  year    = {1932}
}

@article{Majorana1932,
  author  = {E. Majorana},
  title   = {Atomi orientati in campo magnetico variabile},
  journal = {Il Nuovo Cimento},
  volume  = {9},
  number  = {2},
  pages   = {43--50},
  year    = {1932}
}

@article{Shevchenko2010,
  author       = {S. N. Shevchenko and S. Ashhab and F. Nori},
  title        = {Landau--Zener--St\"uckelberg interferometry},
  journal      = {Physics Reports},
  volume       = {492},
  pages        = {1--30},
  year         = {2010}
}

@article{Fox2020,
  title = {Preparing for the quantum revolution: What is the role of higher education?},
  author = {Fox, Michael F. J. and Zwickl, Benjamin M. and Lewandowski, H. J.},
  journal = {Physical Review Physics Education Research},
  volume = {16},
  number = {2},
  pages = {020131},
  year = {2020},
  publisher = {American Physical Society}
}

@article{Gallo1988,
  author  = {Gallo, Giorgio and Pallottino, Stefano},
  title   = {Shortest Path Algorithms},
  journal = {Annals of Operations Research},
  volume  = {13},
  pages   = {1--79},
  year    = {1988},
  doi     = {10.1007/BF02288320}
}

@article{Zermelo1931,
  author  = {Zermelo, Ernst},
  title   = {Über das Navigationsproblem bei ruhender oder veränderlicher Windverteilung},
  journal = {Zeitschrift für Angewandte Mathematik und Mechanik},
  volume  = {11},
  number  = {2},
  pages   = {114--124},
  year    = {1931}
}

@book{McGeoch2014,
  title={Adiabatic Quantum Computation and Quantum Annealing: Theory and Practice},
  author={McGeoch, Catherine C.},
  year={2014},
  publisher={Springer},
  series={Synthesis Lectures on Quantum Computing}
}

@book{Bertsekas2017,
  title={Dynamic Programming and Optimal Control, Vol. I},
  author={Bertsekas, Dimitri P.},
  edition={4th},
  year={2017},
  publisher={Athena Scientific}
}

@article{Glover2019,
  author  = {Fred Glover and Gary Kochenberger and Yu Du},
  title   = {Quantum Bridge Analytics I: A Tutorial on Formulating and Using QUBO Models},
  journal = {4OR},
  volume  = {17},
  number  = {4},
  pages   = {335--371},
  year    = {2019}
}

@article{Feynman1982,
  author  = {Richard P. Feynman},
  title   = {Simulating Physics with Computers},
  journal = {International Journal of Theoretical Physics},
  year    = {1982},
  volume  = {21},
  number  = {6},
  pages   = {467--488}
}

@book{Bellman1965,
  title     = {Dynamic Programming and Modern Control Theory},
  author    = {Bellman, Richard E. and Kalaba, Robert E.},
  year      = {1965},
  publisher = {Academic Press},
  address   = {New York, NY, USA},
  isbn      = {9780120848560},
  note      = {First edition}
}


\end{document}